# Gauge Field Induced Chiral Zero Mode in Five-dimensional Yang Monopole Metamaterials

Shaojie Ma[1,2], Hongwei Jia[4], Yangang Bi[1,5], Shangqiang Ning[1], Fuxin Guan[1], Hongchao Liu[6], Chenjie Wang[1], Shuang Zhang*[1,3]

[1] New Cornerstone Science Laboratory, Department of Physics, The University of Hong Kong; Hong Kong, China.

[2] Department of Optical Science and Engineering, Fudan University, 200433, Shanghai, China

[3] Department of Electrical & Electronic Engineering, The University of Hong Kong; Hong Kong, China.

[4] Department of Physics and Institute for Advanced Study, The Hong Kong University of Science and Technology; Hong Kong, China

[5] State Key Laboratory of Integrated Optoelectronics, College of Electronic Science and Engineering, Jilin University; 2699 Qianjin Street, Changchun 130012, China

[6] Institute of Applied Physics and Materials Engineering, University of Macau, Avenida da Universidade, Taipa, Macao SAR, China

*Corresponding author. Email: shuzhang@hku.hk

**Abstract:**

Owing to the chirality of Weyl nodes characterized by the first Chern number, a Weyl system supports one-way chiral zero modes under a magnetic field, which underlies the celebrated chiral anomaly. As a generalization of Weyl nodes from three-dimensional to five-dimensional physical systems, Yang monopoles are topological singularities carrying nonzero second-order Chern numbers $c_2 = \pm 1$. Here, we couple a Yang monopole with an external gauge field using an inhomogeneous Yang monopole metamaterial, and experimentally demonstrate the existence of a gapless chiral zero mode, where the judiciously designed metallic helical structures and the corresponding effective antisymmetric bianisotropic terms provide the means for controlling gauge fields in a synthetic five-dimensional space. This zeroth mode is found to originate from the coupling between the second Chern singularity and a generalized 4-form gauge field – the wedge product of the magnetic field with itself. This generalization reveals intrinsic connections between physical systems of different dimensions, while a higher dimensional system exhibits much richer supersymmetric structures in Landau level degeneracy due to the internal degrees of freedom. Our study offers the possibility of controlling electromagnetic waves by leveraging the concept of higher-order and higher-dimensional topological phenomena.

Singularities in momentum space, which emerge as low-energy excitations from a multifold degenerate spectrum, play a key role in topological physics [1,2]. For instance, a Weyl semimetal hosts Weyl points (WPs) in the momentum space, which support massless relativistic quasi-particles with quantized Chern numbers $c_1 = \pm 1$ [3-7]. Owing to this topological charge, its quantized Landau band structures under a magnetic field feature a single gapless chiral zero mode (CZM) [8-11], which underlies the celebrated chiral anomaly effect [12,13] and the negative longitudinal magnetoresistance [14].

Generalization of WPs to five-dimensional (5D) space leads to either zero-dimensional Yang monopoles (YMs) [15,16] or two-dimensional linked Weyl surfaces [17-21], both of which possess nontrivial second-order topology with second Chern number $c_2 = \pm 1$. These higher dimensional singularities have been demonstrated in a metamaterial platform constructed by judiciously designed metallic helical structures, with three real momentum dimensions and two bi-anisotropic material parameters as synthetic dimensions [22,23]. 3D Fermi hypersurfaces and 1D Weyl arcs at the 4D boundary of the Yang monopole metamaterial (YMM) were observed, which are key signatures of the nontrivial $c_2$. Since YMs generalize WPs in higher dimensions with second-order topology, a natural question is how they would respond to a gauge field as a result of their nontrivial $c_2$ [24].

A series of recent papers [9,10,25-28] have shown that an artificial gauge field could be applied to a singularity by engineering the individual unit cell to shift the location of the degenerate point spatially. Such designs are excellent platforms for observing Landau levels and CZM induced by the interaction of quasi-particles and artificial external magnetic fields $\vec{B}$. However, limited by the available space-time dimensions, previous demonstrations have been limited to 2D or 3D systems. In this letter, a 5D gauge field $\vec{A}$ is implemented by a judiciously designed inhomogeneous YMM with two synthetic dimensions represented by the antisymmetric bi-anisotropic terms. This gauge field indicates a 4-form background pseudovector field $\vec{T} \propto \vec{B} \wedge \vec{B}$ along the axial z-direction, which matches the order of the differential form of nontrivial nonabelian curvature $\vec{F} \wedge \vec{F}$ [29,30] induced by YM, where $\vec{F}$ is the 2-form Berry curvature and $\wedge$ is the wedge product. This nonabelian curvature is mathematically equivalent to the tensor gauge field $G_{1234}$ discussed in [31]. We for the first time experimentally demonstrate the existence of the generalized gapless CZM induced by the coupling between this pseudovector field $\vec{T}$ and the second Chern singularity in such a higher-dimensional second-order topological system.

We start with the comparison between WP and YM under a gauge field, as shown in Fig. 1(a, b). A typical WP is described by $H_{WP} = \sum_{i=1}^{3} v_i k_i \cdot \sigma_i$, with $\sigma_i$ the Pauli matrices. When coupled with a gauge field $\vec{A} = B_{ij} x_i \hat{e}_j$, by choosing an axis $\hat{e}_3$ along which the pseudovector magnetic field $\vec{B}$ is aligned, the corresponding magnetic field $\vec{B} = B_{12} \hat{e}_3 = B_3 \hat{e}_3$ induces supersymmetric Landau levels in the nonrelativistic squared Hamiltonian:

$$H_{WP,G}^2 = v_3^2 k_3^2 + v_\parallel^2 \cdot [(2n+1) \cdot |B_3| - sgn(v_3) \cdot c_1 B_3 \sigma_z], \tag{1}$$

with $n$ a non-negative integer [32]. The term $v_3^2 k_3^2$ arises from the conserved axial wavevector $k_3$, with $v_3$ the corresponding Fermi velocity. For convenience, we set isotropic horizontal Fermi velocities $v_{i \neq 3} = v_\parallel$. The last term represents the Zeeman term induced by the magnetic field. Except for the zeroth mode, for every eigenstate of Weyl basis $|1\rangle$, there is always another counterpart eigenstate $|2\rangle$ with the same energy and a mode number difference of 1, as shown in Fig. 1(c). At $k_3 = 0$, due to the chiral symmetry $\{H_{WP}, \sigma_3\} = 0$, these supersymmetric structures [32,33] indicate that for a WP under a gauge field there exist symmetric relativistic high order Landau levels and a single CZM. The group velocity of one-way CZM is determined by both the magnetic field $B_3$ and the chirality $c_1$ of WP:

$$\omega_{CZM} = sgn(c_1 B_3) \cdot |v_3| \cdot k_3, \tag{2}$$

as shown in Fig. 1(d) [9-11].

For a YM described by $H_{YM} = \sum_{i=1}^{5} v_i k_i \cdot \Gamma_i$, with $\{\Gamma_i, \Gamma_j\} = 2\delta_{ij}$ satisfying the Clifford algebra, it has a globally doubly degenerate band structure and a fourfold degenerate point [15,16]. The system contains much richer internal structures due to a higher degree of freedom. In the presence of a 5D gauge field $\vec{A}$, in general, there exist ten 2-form magnetic field components $B_{ij}$ and five 4-form pseudovector field components $T_k \propto \varepsilon^{ijkmn} B_{ij} B_{mn}$. By applying a coordinate transformation, one can reduce a uniform 2-form magnetic field to only two components $\{B_{P1}, B_{P2}\}$ individually operating on two separate sets of orthogonal 2-planes, which are both perpendicular to an axis $\hat{e}_3$ along which the pseudovector field $\vec{T}$ is aligned (see section I in SI [34]).

The presence of these fields leads to the following nonrelativistic squared Hamiltonian:

$$H_{YM,G}^2 = \Sigma^2 \pm \sqrt{\Xi + c_2 T_3 \cdot \text{sgn}(v_3) \cdot \Gamma_3}, \tag{3}$$

where $\Sigma^2 = -\sum_j v_j^2 \mathcal{D}_j^2$ with $\mathcal{D}_j \equiv \partial_j - iA_j$, and $\Xi = \frac{1}{2}\sum_{i \neq j} v_i^2 v_j^2 B_{ij}^2$. In a proper Fock space, the two 2-form magnetic field components give rise to supersymmetric Landau levels in the nonrelativistic squared Hamiltonian:

$$H_{YM,G}^2 = v_3^2 k_3^2 + v_\parallel^2 \cdot [|B_{P1}| \cdot (2n_1 + 1) + |B_{P2}| \cdot (2n_2 + 1) - (|B_{P1}|\sigma_3 + |B_{P2}|\sigma_0) \otimes \tau_3], \tag{4}$$

where $\{n_1, n_2\}$ are two non-negative integers, as shown in Fig. 1(e) with $B_{P1} \approx B_{P2}$ [32,35]. The generalized Zeeman term contains two sets of Pauli matrices, where $\sigma$ and $\tau$ operate on the inter-band $|i\rangle$ and intra-band $|\pm\rangle$ spaces, respectively. Due to the $SO(5)$ rotation symmetry [36-38], the supersymmetric structure in the squared YM Hamiltonian possesses much richer degeneracies than its lower dimensional counterpart - the WP, which possesses double degeneracy for all the non-zeroth modes. In the YM system, the degeneracy depends on the energy level, which will be $4N$ at the particular case $|B_{p1}| = |B_{p2}|$, with $N = n_1 + n_2 \neq 0$.

Importantly, there exists an individual CZM, as shown in Fig. 1(f). Interestingly, it is not the 2-form magnetic fields $\vec{B}$, but the 4-form pseudovector field $\vec{T}$ together with $c_2$ of the YM that finally determine the dispersion of CZM, with the direction of the group velocity given by:

$$\omega_{CZM} = \text{sgn}(c_2 T_3) \cdot |v_3| \cdot k_3. \tag{5}$$

Here, the condition guarantees the same Zeeman lift direction in the inter-band space for the two constituent WPs. This generalized CZM is topologically protected by $c_2$. While useful for obtaining the dispersion of the CZM, this squared Hamiltonian cannot fully determine the eigenstates and topological properties. Hence, a more detailed equivalent lattice model in the Fock space is performed [11]. It is verified that the CZM is also protected by an equivalent topological invariant: a pair of opposite nested first Chern number $c_1|_{v_\pm}$ defined on the Wannier sectors [39-42] (see section II-III in SI [34]).

In this work, we focus on this generalized CZM and verify its existence through microwave experiments of inhomogeneous metallic helical YMMs, as shown in Fig. 2. The designed YMM [22] has degenerate electric and magnetic resonances at the plasmonic frequency $\omega_p$. Purely antisymmetric bianisotropic terms $\gamma_{xz} = -\gamma_{zx}$ and $\gamma_{yz} = -\gamma_{zy}$ serve as two synthetic wavevector dimensions $k_4$ and $k_5$, and purely antisymmetric tellegen terms $\varsigma_{ij} = -\varsigma_{ji}$ serve as shifts of three real wavevectors: $\Delta k_k = -\varepsilon_{ijk} \cdot \omega_p \varsigma_{ij}$ (see section IV in SI [34]). A periodic metamaterial [22] just behaves like a 5D YM with 4-fold degeneracy located at $[\vec{K}_{YM}, \omega_p]$, in which the space of Clifford operators $\Gamma_i$ is spanned by two degenerate longitudinal plasma modes $\{|E_z\rangle, |H_z\rangle\}$ with flat dispersion and two transverse electromagnetic modes $\{|E_x\rangle, |E_y\rangle\}$ (brown lines in Fig. 2(d)).

For an inhomogeneous metamaterial, following the rules of minimal coupling $\partial_j \mapsto \mathcal{D}_j \equiv \partial_j - iA_j$ in the usual covariant derivative argument, a vector gauge field $\vec{A}$ can be viewed as a space-dependent shift of the YM locations $\vec{K}_{YM}(\vec{r})$ in the 5D momentum space, with magnetic field $B_{ij} = i \cdot [\mathcal{D}_i, \mathcal{D}_j] = \partial_i A_j - \partial_j A_i$ caused by this spatial shift [43]. Without loss of generality, we choose the axial direction along the $z$-direction. Therefore, the inhomogeneous YMM slow-varying in $xy$-plane can introduce an arbitrary nontrivial gauge field $\vec{A}(\vec{r}) = A_i(x,y) \cdot \hat{e}_i$ by designing the space-dependent magneto-electric tensor [44]. Compared with the homogeneous system, this inhomogeneous system not only introduces the magnetic fields $B_{ij}$ that can couple with the first-order topological singularity, but also contains a nontrivial 4-form field component $T_3 = \frac{1}{4} v_\parallel^4 \cdot \vec{B} \wedge \vec{B} \cdot \hat{e}_3 = 2v_\parallel^4 \cdot (B_{15}B_{24} - B_{14}B_{25} + B_{12}B_{45})$ along the $z$-direction, which interacts with the second-order topological singularity of the YM and induces the generalized CZM in 5D photonic YMM.

Figure 2(a) shows a schematic diagram for a specific inhomogeneous YMM with only nonzero $B_{15}$, $B_{24}$ and all other $B_{ij}$=0. The shift of the YM only occurs in the synthetic dimensions by varying the bianisotropy terms: $A_4 = -\omega_p \gamma_{xz}(\vec{r}) = B_{24} \cdot y$ and $A_5 =$

$-\omega_p \gamma_{yz}(\vec{r}) = B_{15} \cdot x$, which correspond to two individual space-dependent mass terms in three real dimensions [45,46]. Note that such a configuration does not require any tellegen materials for observing the CZM. The space-dependent bianisotropy distribution is realized by a set of rotated metallic helical units, as shown in Fig. 2(b, c). In each unit, four precisely adjusted helical structures combined with their mirror counterparts are collectively rotated to the angles $\Phi_{1\mapsto 4} = \psi_{45} + [+\delta_{45}, +\delta_{45}+90°, -\delta_{45}+180°, -\delta_{45}+270°]$, which can realize purely antisymmetric bianisotropic terms satisfying $\gamma_{xz} + i\gamma_{yz} \propto \sin\delta_{45} \cdot \exp[i(\psi_{45} + 45°)]$ [22]. Therefore, $\vec{A}(\vec{r})$ can be realized in this inhomogeneous photonic metamaterial through an appropriate spatial distribution of rotation angles $[\delta_{45}, \psi_{45}]$. In the experimental demonstration, we design $\sin\delta_{45}$ to be linearly varying with radius, which varies from 0 to 1 through 20 units, and a space-dependent phase distribution $\psi_{45} = atan2(x, y) - 45°$. This inhomogeneous YMM contains a uniform effective magnetic flux density $B_{15} = B_{24} \approx -1210 \text{m}^{-2}$ generated by the spatially shifted $\vec{A}(\vec{r})$ and a uniform 4-form pseudovector field $T_3 \approx (0.011\omega_p)^4$ along the $z$-direction, which opens up a sufficiently large enough bandgap of approximately 0.39GHz (See section IV in SI [34]).

The local (left panel) and global (right panel) dispersions of this metamaterial are shown in Fig. 2(d). Locally, a nonzero angle $\delta_{45}$ behaves like an effective mass, which opens a bandgap and constructs two pairs of degenerate bands near YM. Meanwhile, globally this inhomogeneous metamaterial supports a single $E_x$-polarized confined state near the original YM, which is the CZM induced by the nontrivial field $T_3$ and protected by $c_2$. This eigenstate corresponds to a localized zero-order Hermite-Gauss field distribution, as shown in Fig. 2(e). Thus, a polarization-dependent dispersion spectrum can be measured to verify this CZM.

The sample is constructed by stacking up 160 printed circuit board layers (40 unit cells along the $z$-direction), as shown in Fig. 3(a). A linearly polarized wave is launched by a horn antenna located below the center of the bottom layer, while the field distribution inside the vertical slits is detected by the near-field scanning of a monopole antenna aligned to a co-polarization direction. Fig. 3(b) shows the simulated and measured co-polarization field distribution at the plasma frequency around 14.66GHz in different polarization setups. The

$E_x$-polarized field can propagate through the metamaterial, while the $E_y$-polarized field decays rapidly along the $z$ direction, agreeing with our theoretical prediction that this inhomogeneous metamaterial supports a single $E_x$-polarized CZM, but behaves as a bandgap for the $E_y$-polarized excitation. A significant contrast about 30dB between different polarizations is observed in the measured transmitted power near the plasma frequency, as shown in Fig. 3(e, f). The dispersion spectra of the two polarizations along $k_z$, obtained through Fourier transformation of the field patterns, show a significant difference near the plasma frequency from about 14.55GHz to 14.71GHz, both in simulation and in the experiment, as shown from the comparison between Fig. 3(c, d). Such difference in dispersion spectrum between the two polarizations is consistent across a series of measurements at different in-plane locations. In comparison, the dispersion spectrum and transmitted power are both nearly polarization-independent at frequencies away from the plasma frequency (See section V in SI [34] for experimental details). This contrast provides direct evidence for the presence of polarized zeroth mode near YM.

In summary, we have explored the interaction of higher-order topological singularities with a gauge field in a 5D system, and we experimentally demonstrated the existence of CZM by employing an inhomogeneous metamaterial platform. Under a gauge field, the YMs with nontrivial $c_2$ provide a much richer Landau structure than its 3D counterpart, due to the interplay between 2-form magnetic fields $\vec{B}$ and 4-form pseudovector fields $\vec{T}$. Interestingly, the formation of the CZM directly results from the interaction between $c_2$ and the pseudovector $\vec{T}$ field, which serves as a new manifestation of the intriguing topological properties of YM. Our work provides new approaches for electromagnetic control by exploiting the combination of higher-dimensional topology and artificially engineered gauge fields.

**Acknowledgments:**

We would like to thank Prof. T.T. Luu for his help in sample preparation. This work was supported by the Research Grants Council of Hong Kong (AoE/P-701/20, 17309021, 16307621), and the start-up funding of Fudan University (JIH1232082Y)

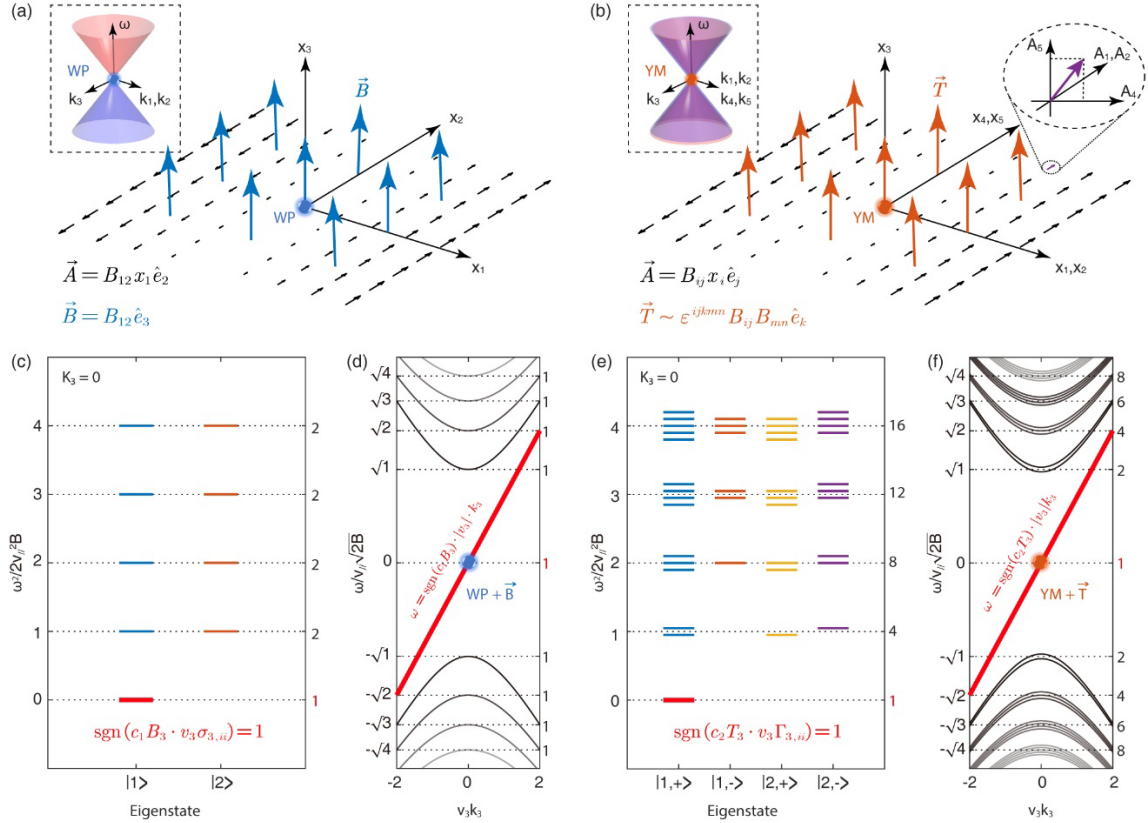

**Fig. 1. Illustration of Landau Levels and CZM in WP and YM under a Gauge Field.** (a) A WP in 3D space under an external magnetic field $\vec{B}$. (b) The counterpart of (a) in 5D space - a YM under an external 4-form pseudovector $\vec{T}$ field. The upper-left inset in (a, b) shows the linear dispersion spectrum near the singularity, and the upper-right inset in (b) shows a 5D gauge field $\vec{A}$. (c) The supersymmetric Landau levels corresponding to the nonrelativistic squared WP Hamiltonian. (d) The dispersion spectrum of the relativistic WP Hamiltonian, with the red straight line representing the CZM. (e, f) The counterpart of (c, d) for YM, with two effective magnetic field components $B_{P1} \approx B_{P2}$. The numbers on the right vertical axis indicate the degeneracy of each set of Landau levels if $|B_{P1}| = |B_{P2}|$.

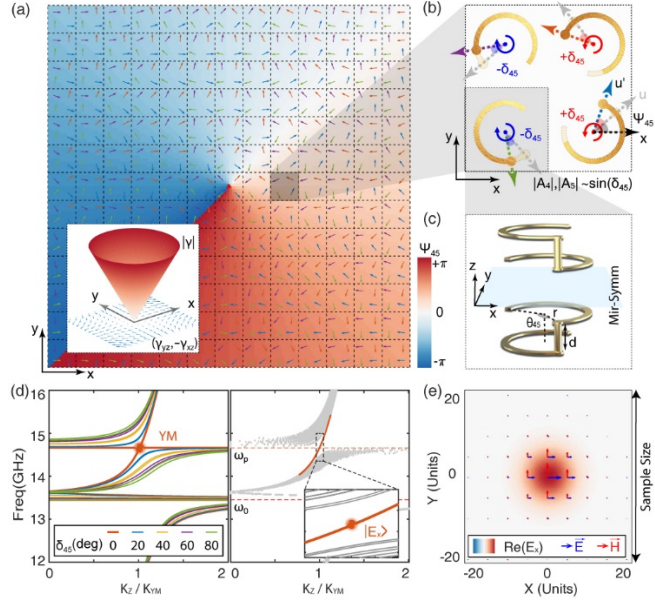

**Fig. 2. Illustration of Inhomogeneous Yang Metamaterial.** (a-c) The configuration of an inhomogeneous YMM with $A_4 = B_{24} \cdot y$ and $A_5 = B_{15} \cdot x$. The inset in (a) represents the magnitude and angle distribution of the bianisotropy vectors, and the color-map labels the spatial distribution of $\psi_{45}$. (b, c) The top view and side view of the metallic helices, respectively. The spatial distribution of the two angles $\delta_{45}$ and $\psi_{45}$ is precisely designed to achieve an arbitrary gauge field distribution. (d) The local (left) and global (right) dispersion of the designed inhomogeneous metamaterial along the axial direction $k_z$. The brown line represents the dispersion of the original YM (left) and CZM (right). (e) The field distribution of CZM at the original YM location.

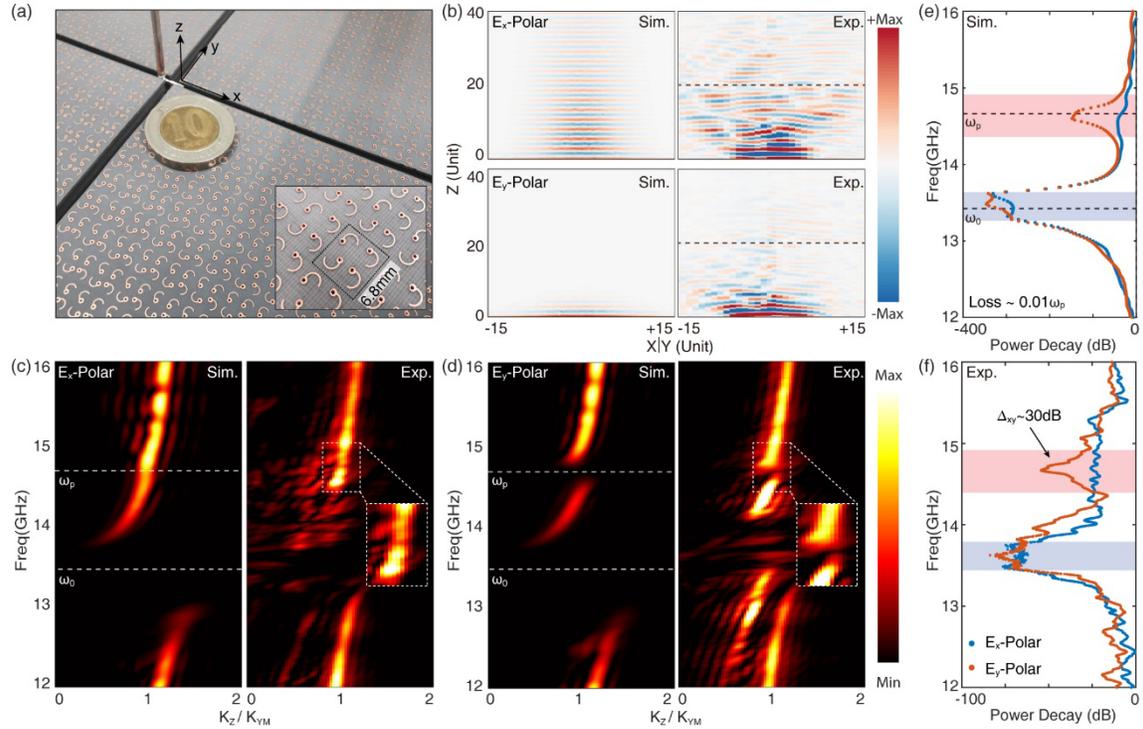

**Fig. 3. Experimental Observation of the Polarization-dependent Dispersion of the 5D CZM.** (a) Photograph of the top surface of the sample, fabricated with printed circuit board technology, with one unit cell indicated by the black square. Two 3mm-wide vertical slits are cut through the center of the sample to measure the field distribution inside the metamaterial. (b) The simulated (left) and measured (right) electric field distribution in the slits by different polarized excitations at the plasma frequency. The data $z \in (0,20)$ units and $z \in (20,40)$ units are from two independent measurements. (c, d) The simulated (left) and measured (right) dispersion spectra by (c) $E_x$-polarized and (d) $E_y$-polarized excitations. In both cases, the direction of the probe antenna is aligned to the polarization of the wave launched by the horn antenna, and $E_x/E_y$ polarization is detected in the XZ/YZ slit, respectively. (e, f) The (e) simulated and (f) measured transmitted power for different polarizations obtained from integration along the corresponding slit. The $K_{YM}$ positions of the experimental data are normalized individually.

# Supplementary Materials:

# Gauge Field Induced Chiral Zero Mode in Five-dimensional Yang Monopole Metamaterials

Shaojie Ma[1,2], Hongwei Jia[4], Yangang Bi[1,5], Shangqiang Ning[1], Fuxin Guan[1], Hongchao Liu[6], Chenjie Wang[1], Shuang Zhang*[1,3]

*Corresponding author. Email: shuzhang@hku.hk

**This file includes:**

1. Landau Levels in WP and YM systems

2. Eigenstates of Landau Levels in YM systems

3. Discussion on Second Chern Number and Nested Chern Number

4. Discussion on Effective Medium Model and Simulation Model

5. Discussion on Experimental Measurements

**References**

**Figs. S1 to S14**

## 1.   Landau Levels in WP and YM systems

We will compare the intrinsic similarities and differences between Weyl point (WP) and Yang monopole (YM) in terms of the dispersion spectra of their Landau levels, as shown in Fig. 1.

For simplicity, we only focus on a minimal Hamiltonian $H_{WP} = \sum_{i=1}^{3} v_i k_i \sigma_i$ for WP and $H_{YM} = \sum_{i=1}^{5} v_i k_i \Gamma_i$ for YM. The topologies of two singularities are determined by the integer first and second Chern numbers, respectively. These topological invariants are both determined by the signs of the product of all Fermi velocities in these minimal Hamiltonians: $c_{1(2)} = sgn(\prod_i v_i) = \pm 1$.

The set of three Pauli matrices and the set of five Dirac Gamma matrices both satisfy the same Clifford algebra, i.e., $\{\sigma_i, \sigma_j\} = 2\delta_{ij}$ and $\{\Gamma_i, \Gamma_j\} = 2\delta_{ij}$, with $\delta_{ij}$ being the Kronecker delta function, which guarantees the linear dispersion spectrum near the singularity $\omega = \pm |\vec{v} \cdot \vec{k}|$:

$$\omega^2 \Phi = H^2 \Phi = \begin{cases} \dfrac{1}{2}\sum_{i,j}\{v_i k_i \sigma_i, v_j k_j \sigma_j\} \cdot \Phi = \sum_{i=1}^{3} v_i^2 k_i^2 \cdot \Phi \\ \dfrac{1}{2}\sum_{i,j}\{v_i k_i \Gamma_i, v_j k_j \Gamma_j\} \cdot \Phi = \sum_{i=1}^{5} v_i^2 k_i^2 \cdot \Phi \end{cases} \Rightarrow \omega = \pm|\vec{v} \cdot \vec{k}|, \text{(S1)}$$

as shown in the inset of Fig.1(a, b). Furthermore, YM also satisfies an additional **PT-**symmetry (space inversion and time-reversal symmetry) with $(PT)^2 = -1$, which indicates a fourfold singularity formed by two pairs of degenerate linear bands.

When coupled with a background field through a gauge field $\vec{A} = B_{ij} x_i \hat{e}_j$ as shown in Fig. 1(a, b), non-commutative algebra (Lie algebra) will lead to Landau levels in both types of singularities. Notably, both singularities support the chiral zeroth Landau mode, but they are protected by different Chern numbers and induced by different fields. We will discuss these Landau levels in detail.

## 2.   Dispersion spectra of Landau Levels and Chiral zero mode in WP

A WP can support a chiral zero mode (CZM), with group velocity determined by both the direction of the 2-form magnetic field $\vec{B}$ and the chirality (first Chern number $c_1$) of

WP. This is a well-known result. Here, we will give a simple derivation based on the squared Hamiltonian and verify that the CZM and the associated Landau levels are induced by the non-commutative algebra $[\sigma_i, \sigma_j] = 2i\varepsilon_{ijk}\sigma_k$.

Without loss of generality, we consider a background magnetic field $\vec{B} = B_{12}\hat{e}_3 = B_3\hat{e}_3$ along the z-direction. The Hamiltonian of WP will couple with a gauge field $\vec{A} = B_{ij}x_i\hat{e}_j$. This field $\vec{A}$ is gauge dependent. For example, we can choose a symmetric gauge field $\vec{A}_{SG} = \frac{1}{2} \cdot [-B_3 x_2, B_3 x_1, 0]$, a Landau gauge field $\vec{A}_{LG} = [0, B_3 x, 0]$, or any other gauge fields $\vec{A}$ satisfying $\vec{B} = \nabla \times \vec{A}$. In all these cases, the dispersion spectra of the Hamiltonian under a gauge field will remain the same, i.e., they are gauge-independent.

Here, we choose a symmetric gauge field $\vec{A}_{SG}$ for illustration. The Hamiltonian for WP is given by:

$$H_{WP,G} = v_1\left(-i\partial_1 + \frac{1}{2}B_3 x_2\right)\sigma_1 + v_2\left(-i\partial_2 - \frac{1}{2}B_3 x_1\right)\sigma_2 + v_3 k_3 \sigma_3. \quad (S2)$$

In this Hamiltonian, due to the presence of $x_1$ and $x_2$, the momentum wavevectors $k_1$ and $k_2$ are not good quantum numbers and therefore should be replaced by $k_j \mapsto -i\mathcal{D}_j \equiv -i\partial_j - A_j$ following the rule of minimal coupling based on the usual covariant derivative argument, with the magnetic field satisfying $B_3 = i \cdot [\mathcal{D}_1, \mathcal{D}_2]$. To solve its dispersion, a traditional method is to solve a non-relativistic squared Hamiltonian:

$$H_{WP,G}^2 = \sum_i (-iv_i \mathcal{D}_i)^2 + v_1 v_2 \cdot [-i\mathcal{D}_1, -i\mathcal{D}_2] \cdot \sigma_1 \sigma_2 = \Sigma^2 - v_\parallel^2 \cdot \text{sgn}(v_3) c_1 B_3 \cdot \sigma_3.$$

(S3)

Here, we have set the system to be isotropic in the plane perpendicular to the applied magnetic field, i.e., $v_1 = v_2 = v_\parallel$, and we have used the conditions $c_1 = sgn(v_1 v_2 v_3) = \pm 1$, and the non-commutative algebra $[\sigma_i, \sigma_j] = 2i\varepsilon_{ijk}\sigma_k$.

Define $\alpha = \sqrt{|B_3|}$. The summation term $\Sigma^2$ can be directly solved when considering the similarity of its mathematical form to quantum harmonic oscillators:

$$\Sigma^2 = v_3{}^2 k_3{}^2 + v_\|{}^2 |B_3| \cdot \left[ \left( \frac{-i\partial_1}{\alpha} + \mathrm{sgn}(B_3) \cdot \frac{1}{2}\alpha x_2 \right)^2 + \left( \frac{-i\partial_2}{\alpha} - \mathrm{sgn}(B_3) \cdot \frac{1}{2}\alpha x_1 \right)^2 \right]$$

$$= v_3{}^2 k_z{}^2 + v_\|{}^2 |B_3| \cdot \left( \frac{\hat{a} - \hat{a}^\dagger}{\sqrt{2}i} \right)^2 + v_\|{}^2 |B_3| \cdot \left( -\mathrm{sgn}(B_3) \cdot \frac{\hat{a} + \hat{a}^\dagger}{\sqrt{2}} \right)^2$$

$$= v_3{}^2 k_3{}^2 + v_\|{}^2 |B_3| \cdot (2\hat{n} + 1)$$

(S4)

The particle number operator $\hat{n} = \hat{a}^\dagger \hat{a}$ represents a non-negative integer on the Fock basis of an effective quantum harmonic oscillator model. Here, the creation and annihilation operators satisfy the canonical commutation relation $[\hat{a}, \hat{a}^\dagger] = 1$, with:

$$\begin{aligned}
\hat{a} &= \frac{1}{\sqrt{2}} \left[ \frac{1}{\alpha}(\partial_1 + \mathrm{sgn}(B_3) \cdot i\partial_2) + \frac{1}{2}\alpha \cdot (x_1 + \mathrm{sgn}(B_3) \cdot ix_2) \right] \\
\hat{a}^\dagger &= \frac{1}{\sqrt{2}} \left[ -\frac{1}{\alpha}(\partial_1 - \mathrm{sgn}(B_3) \cdot i\partial_2) + \frac{1}{2}\alpha \cdot (x_1 - \mathrm{sgn}(B_3) \cdot ix_2) \right]
\end{aligned}$$

(S5)

Therefore, this magnetic field $\vec{B} = B_{12}\hat{e}_3 = B_3\hat{e}_3$ leads to supersymmetric Landau levels in the non-relativistic squared WP Hamiltonian:

$$H_{WP,G}{}^2 = v_3{}^2 k_3{}^2 + v_\|{}^2 \cdot [|B_3| \cdot (2n+1) - \mathrm{sgn}(v_3) \cdot c_1 B_3 \cdot \sigma_3]. \quad (S6)$$

Except for the zeroth mode, for every eigenstate of Weyl basis $|1\rangle$, there is always another counterpart eigenstate $|2\rangle$ with the same energy and a mode number difference of 1, as shown in Fig. 1(c). In the above equation, the last term represents the Zeeman term induced by the magnetic field, which is derived from the non-commutative algebra $[\sigma_i, \sigma_j] = 2i\varepsilon_{ijk}\sigma_k$. This Zeeman term is directly related to the magnetic field $\vec{B}$, which is gauge-independent.

For the zeroth Landau level with $n = 0$, the Zeeman term exactly cancels the zero-point energy induced by the effective quantum harmonic oscillators, which means:

$$\mathrm{sgn}(v_3) \cdot c_1 B_3 \cdot \sigma_{3,ii} = |B_3| \Rightarrow \mathrm{sgn}(v_3 c_1 B_3 \cdot \sigma_{3,ii}) = 1. \quad (S7)$$

Here, $\sigma_{3,ii}$ is the diagonal element of the Pauli matrix $\sigma_3$, and the condition $\sigma_{3,ii} = 1$ or $\sigma_{3,ii} = -1$ in Eq. S7 specifies that the eigenstate of the CZM is $|1\rangle$ or $|2\rangle$, respectively.

This individual zero mode, shown by the red line in Fig. 1(d), has a chiral group velocity determined by both the direction of the magnetic field $B_3$ and the first Chern number $c_1$:

$$\omega_{CZM} = \text{sgn}\,(v_3 c_1 B_3 \sigma_{3,ii}) \cdot \sigma_{3,ii} v_3 k_3 = \text{sgn}\,(c_1 B_3) \cdot |v_3| k_3. \tag{S8}$$

Furthermore, by switching to a Landau gauge, the only changed term is the summation term, with $\Sigma_{LG}^2 = v_3^2 k_3^2 + v_\parallel^2 k_1^2 + v_\parallel^2 B_3 \cdot (k_2/B_3 - x_1)^2$. However, with a changed ladder operator $\hat{a} = \frac{1}{\sqrt{2}}\left[\frac{1}{\alpha}\partial_1 + \alpha(x - k_2/B_3)\right]$, the summation can be the same: $\Sigma_{LG}^2 = v_3^2 k_3 + v_\parallel^2 |B_3| \cdot (2n+1)$. This summation is indeed gauge-independent. It implies that the entire dispersion spectrum is gauge-independent, which only depends on the magnetic field $\vec{B}$, but not the specific gauge field $\vec{A}$.

It is worth mentioning that in both cases, one partial derivative operator will be redundant, e.g., $k_2$ in the Landau gauge or one of the effective $k_\pm = k_1 \pm i k_2$ in the symmetric gauge. Thus, we can artificially select an arbitrary value of $k_y$ or $k_\pm$, which only affects the field distribution of eigenstate but not the dispersion at all.

### 3. Dispersion spectra of Landau Levels and Chiral zero mode in YM

A YM in 5D space can also support CZM, but its group velocity is determined by both the direction of a 4-form pseudovector field and the second Chern number of YM, which reveals intrinsic connections between the order of topology and the differential forms of the responded external gauge field. The Landau levels contain much richer internal structures than those of WP due to the non-closed non-commutative algebra $[\Gamma_i, \Gamma_j] \notin c_k \Gamma_k$ for the five Gamma matrices.

Here we consider a general case in which the Hamiltonian for YM couples with a gauge field $\vec{A} = A_i \hat{e}_i$ in 5D space, which in general can lead to a 2-form magnetic field with 10 components $B_{ij}\,(i \neq j)$, with $B_{ij} = -B_{ji} = i[\mathcal{D}_i, \mathcal{D}_j] = (\partial_i A_j - \partial_j A_i) - i[A_i, A_j]$ and $\mathcal{D}_j \equiv \partial_j - i A_j$. Here, $[A_i, A_j] = 0$ for the Abelian U(1) gauge field representing an electromagnetic background field. In this article, we will focus on this particular situation. In the future, more general cases, such as a non-Abelian SU(2) gauge field representing the Yang-Mills background field, can be discussed. The Hamiltonian can be expressed as:

$$H_{YM,G} = -i\sum_{j=1}^{5} v_j \mathcal{D}_j \cdot \Gamma_j = \sum_{j=1}^{5} v_j(-i\partial_j - A_j)\cdot \Gamma_j. \tag{S9}$$

Similarly, we will solve a non-relativistic squared Hamiltonian to analyze its dispersion spectrum:

$$H_{YM,G}^2 = -\sum_i v_i^2 \mathcal{D}_i^2 + \frac{1}{2}\sum_{i\neq j} v_i v_j \cdot i[\mathcal{D}_i, \mathcal{D}_j] \cdot i\Gamma_i \Gamma_j = \Sigma^2 + \frac{1}{2}\sum_{i\neq j} v_{ij} B_{ij} \Gamma_{ij}.$$

$$\tag{S10}$$

Here, the definitions $v_{ij} = v_i v_j$, $\Gamma_{ij} = \frac{i}{2}[\Gamma_i, \Gamma_j] = i\Gamma_i \Gamma_j$ when $i \neq j$, and $\Gamma_{ii} = 0$. The summation term $\Sigma^2 = -\sum_i v_i^2 \mathcal{D}_i^2$ represents the direct summation of squares of all effective wavevectors, which is gauge-independent. The coexistence of the space operator and the momentum operators in the squared Hamiltonian leads to a mathematic form equivalent to effective quantum harmonic oscillators, and consequently the existence of zero-point energy.

This squared Hamiltonian has ten $\Gamma_{ij}$ terms coupled with the ten 2-form magnetic field components $B_{ij}$. When all the magnetic field components are *homogeneous*, i.e., $\partial_i B_{jk} = 0$, an additional squaring operation can help to diagonalize the squared Hamiltonian, that is:

$$\begin{aligned} H_{YM,G}^4 &= \left(\Sigma^2 + \frac{1}{2}\sum_{i\neq j} v_{ij} B_{ij} \cdot \Gamma_{ij}\right)\left(\Sigma^2 + \frac{1}{2}\sum_{i\neq j} v_{ij} B_{ij} \cdot \Gamma_{ij}\right) \\ &= \Sigma^4 + 2\Sigma^2 \cdot \frac{1}{2}\sum_{i\neq j} v_{ij} B_{ij} \Gamma_{ij} + \frac{1}{4}\sum_{i\neq j} v_{ij} B_{ij} \Gamma_{ij} \cdot \sum_{m\neq n} v_{mn} B_{mn} \Gamma_{mn} \\ &= \Sigma^4 + 2\Sigma^2 \cdot (H_{YM,G}^2 - \Sigma^2) + \frac{1}{8}\sum_{i\neq j}\sum_{m\neq n}\{v_{ij} B_{ij}\Gamma_{ij}, v_{mn} B_{mn} \Gamma_{mn}\} \\ &= (2\Sigma^2 \cdot H_{YM,G}^2 - \Sigma^4) + \frac{1}{2}\sum_{i\neq j} v_{ij}^2 B_{ij}^2 + \frac{1}{4}\sum_{i,j,m,n} \varepsilon^{ijkmn} v_{ij} v_{mn} B_{ij} B_{mn} \Gamma_k \end{aligned} \tag{S11}$$

Here, the anti-commutation relation of $\Gamma_{ij}$ satisfies:

$$\{\Gamma_{ij}, \Gamma_{mn}\} = 2I \cdot (\delta_{im}\delta_{jn} - \delta_{in}\delta_{jm}) + 2\Gamma_k \cdot \varepsilon^{ijkmn}. \tag{S12}$$

Considering $c_2 = sgn(v_1 v_2 v_3 v_4 v_5) = \pm 1$, we can define the following two expressions:

$$\Xi = \frac{1}{2} \sum_{i \neq j} v_{ij}{}^2 B_{ij}{}^2 = \begin{pmatrix} v_{12}^2 B_{12}^2 + v_{13}^2 B_{13}^2 + v_{14}^2 B_{14}^2 + v_{15}^2 B_{15}^2 + v_{23}^2 B_{23}^2 \\ + v_{24}^2 B_{24}^2 + v_{25}^2 B_{25}^2 + v_{34}^2 B_{34}^2 + v_{35}^2 B_{35}^2 + v_{45}^2 B_{45}^2 \end{pmatrix}, \quad (S13.1)$$

$$\begin{aligned}
\hat{\mathbb{T}} &= \frac{1}{4} \sum_{i,j,m,n} c_2 \cdot \varepsilon^{ijkmn} v_{ij} v_{mn} B_{ij} B_{mn} \cdot \Gamma_k \\
&= \frac{1}{4} \sum_{i,j,m,n} \Pi_v \operatorname{sgn}(v_k)/|v_k| \cdot \varepsilon^{ijkmn} B_{ij} B_{mn} \cdot \Gamma_k \\
&= \begin{cases}
+ 2\operatorname{sgn}(v_1) \cdot \Pi_v/|v_1| \cdot (+B_{23} B_{45} - B_{24} B_{35} + B_{25} B_{34}) \cdot \Gamma_1 \\
+ 2\operatorname{sgn}(v_2) \cdot \Pi_v/|v_2| \cdot (-B_{34} B_{15} - B_{13} B_{45} + B_{14} B_{35}) \cdot \Gamma_2 \\
+ 2\operatorname{sgn}(v_3) \cdot \Pi_v/|v_3| \cdot (+B_{12} B_{45} - B_{14} B_{25} + B_{15} B_{24}) \cdot \Gamma_3 \\
+ 2\operatorname{sgn}(v_4) \cdot \Pi_v/|v_4| \cdot (-B_{15} B_{23} - B_{12} B_{35} + B_{13} B_{25}) \cdot \Gamma_4 \\
+ 2\operatorname{sgn}(v_5) \cdot \Pi_v/|v_5| \cdot (+B_{12} B_{34} - B_{13} B_{24} + B_{14} B_{23}) \cdot \Gamma_5
\end{cases}
\end{aligned} \quad (S13.2)$$

Here, $\Pi_v = |v_1 v_2 v_3 v_4 v_5|$. The expression $\Xi$ represents the summation of squares of all magnetic field components, and $\hat{\mathbb{T}}$ represents a 4-form tensor field proportional to $\vec{B} \wedge \vec{B}$, with $\wedge$ the wedge product.

The non-relativistic quartic eigenequation $H_{YM,G}^4 \Phi = \omega^4 \Phi$ and non-relativistic quadratic eigenequation $H_{YM,G}^2 \Phi = \omega^2 \Phi$ can be solved directly:

$$\begin{aligned}
H_{YM,G}^4 \Phi &= \left[ (-\Sigma^4 + 2\Sigma^2 \cdot H_{YM,G}^2 + \Xi) \cdot I + c_2 \hat{\mathbb{T}} \right] \cdot \Phi = \omega^4 \Phi \\
&\Rightarrow c_2 \hat{\mathbb{T}} \cdot \Phi = \left[ (\omega^2 - \Sigma^2)^2 - \Xi \right] \cdot \Phi \\
&\Rightarrow H_{YM,G}^2 = \Sigma^2 \pm \sqrt{\Xi + c_2 \hat{\mathbb{T}}} = \Sigma^2 \pm \sqrt{\Xi \pm c_2 |\vec{T}|}
\end{aligned} \quad (S14)$$

The last step just follows Eq. S1 by noting that $\hat{\mathbb{T}} = \sum_i \operatorname{sgn}(v_i) \cdot T_i \cdot \Gamma_i$, with $\vec{T} = \operatorname{sgn}(v_i) \cdot T_i \cdot \hat{e}_i$ an effective 5D pseudovector vector. This 4-form field $\vec{T}$ behaves just like the pseudovector magnetic field in a 3D system, which is indeed a second-order pseudovector field, as shown in Fig. S1 (b). The differential forms of this field $\vec{T}$ just corresponds to the order of topology, and this pseudovector field $\vec{T}$ can couple to a second topological system with $c_2 = \pm 1$ and induce a CZM, in a similar way as the interaction between a magnetic field $\vec{B}$ and a WP with $c_1 = \pm 1$.

Without loss of generality, we can choose the direction of $\vec{T}$ as the axial direction and directly define a coordinate system with the axial direction along $x_3$, and choose a basis satisfying $\Gamma_3 = \text{diag}([1,1,-1,-1])$. This setup is the counterpart of the WP system with $\vec{B} = B_{12}\hat{e}_3 = B_3\hat{e}_3$ and $\sigma_3 = \text{diag}([1,-1])$. For convenience, we also set isotropic horizontal Fermi velocities $v_{i\neq 3} = v_\parallel$. In this specific coordinate system, we have $B_{i3} = 0$, $A_3 = 0$, $\Xi = v_\parallel^2 \cdot (B_{12}^2 + B_{14}^2 + B_{15}^2 + B_{24}^2 + B_{25}^2 + B_{45}^2)$, and $\widehat{\mathbb{T}} = \text{sgn}(v_3) \cdot T_3\Gamma_3$ with $T_3 = v_\parallel^4/4 \cdot \vec{B}\wedge\vec{B} \cdot \hat{e}_3 = 2v_\parallel^4 \cdot (B_{12}B_{45} - B_{14}B_{25} + B_{15}B_{24})$, as shown in Fig. S1(c). Thus, we have:

$$H_{YM,G}^2 = \Sigma^2 \pm \sqrt{\Xi + c_2 T_3 \cdot \text{sgn}(v_3)\Gamma_3}. \tag{S15}$$

The last square root term is the generalized Zeeman term in a higher dimensional system, which depends on both $\Xi$ and $T_3$. This term can cancel the zero-point energy induced by the effective quantum harmonic oscillators in $\Sigma^2$ and induce a generalized CZM.

Considering that the summation $\Sigma^2$ is gauge-independent, we can choose an arbitrary gauge field $\vec{A}$ satisfying $B_{ij} = i[\mathcal{D}_i,\mathcal{D}_j] = (\partial_i A_j - \partial_j A_i)$ to calculate its value. Here, we choose a symmetric gauge field $\vec{A}_{SG}$ for illustration:

$$\begin{bmatrix} A_1 \\ A_2 \\ A_4 \\ A_5 \end{bmatrix} = S \cdot \begin{bmatrix} x_1 \\ x_2 \\ x_4 \\ x_5 \end{bmatrix} = \frac{1}{2}\begin{bmatrix} 0 & -B_{12} & -B_{14} & -B_{15} \\ B_{12} & 0 & -B_{24} & -B_{25} \\ B_{14} & B_{24} & 0 & -B_{45} \\ B_{15} & B_{25} & B_{45} & 0 \end{bmatrix} \cdot \begin{bmatrix} x_1 \\ x_2 \\ x_4 \\ x_5 \end{bmatrix}. \tag{S16}$$

The matrix $S$ is an antisymmetric square matrix satisfying $S^T = -S$. From a geometrical point of view, this antisymmetric matrix represents a $SO(4)$ rotation about a fixed point $(0,0,0,0)$ in four-dimensional Euclidean space [1,2]. A Four-dimensional rotation can be derived from Rodrigues' rotation formula or the Cayley formula, with $S$ uniquely decomposed as:

$$S = \theta_1 \cdot S_1 + \theta_2 \cdot S_2. \tag{S17}$$

$S_i$ is also an antisymmetric matrix satisfying $S_i^T = -S_i$, $S_1 S_2 = 0$ and $S_i^3 = -S_i$. The angles $\theta_1$ and $\theta_2$ are obtained from the eigenvalues of the antisymmetric matrix $S$, with

$\Lambda = \text{eig}(S) = \text{diag}([i\theta_1, -i\theta_1, i\theta_2, -i\theta_2])$. Without loss of generality, we define $\theta_1 \geq \theta_2 \geq 0$. These angles $\theta_i$ can be expressed as:

$$\theta_1^2 + \theta_2^2 = \frac{1}{4} \cdot \Xi \,\,\&\,\, \theta_1 \theta_2 = \frac{1}{8}|T_3|$$

$$\theta_1 = \frac{|B_{P1}|}{2} = \frac{\sqrt{2}}{4} \cdot \sqrt{\Xi + \sqrt{\Xi^2 - T_3^2}} \,\,\&\,\, \theta_2 = \frac{|B_{P2}|}{2} = \frac{\sqrt{2}}{4} \cdot \sqrt{\Xi - \sqrt{\Xi^2 - T_3^2}}.$$

(S18)

The two angles correspond to two effective magnetic field components $|B_{P1}| = 2\theta_1$ and $|B_{P2}| = 2\theta_2$, as shown in Fig. S1(d). From a geometrical point of view, this decomposition separates one pair of orthogonal 2-planes: $P_1$ and $P_2$, with $S_i$ operating only on one of these planes and producing an ordinary rotation $\theta_i$. Mathematically, this result depends on the fact that the pair of invariant planes of a commutative subgroup of $SO(4)$ is isomorphic to $SO(2) \times SO(2)$.

Therefore, we can consider an orthogonal transformation for the 4D Euclidean space perpendicular to $x_3$, with $S$ directly operating on two separated orthogonal 2-planes. This orthogonal transformation can be constructed through the eigenstates of two projection operators $P_i = -S_i^2$ satisfying $P_i S = \theta_i S_i$:

$$P_i V_{i\pm} = V_{i\pm}, \text{ with } V_0 = [V_{1+}, V_{1-}, V_{2+}, V_{2-}] \text{ and } V_0^T V_0 = I. \quad (S19)$$

These eigenstates satisfy: $P_i V_{j\pm} = \delta_{ij} V_{j\pm}$. Meanwhile, we have $S_i V_{j\pm} = \pm \delta_{ij} V_{j\mp}$ or $S_i V_{j\pm} = \mp \delta_{ij} V_{j\mp}$ based on the orientations of the 2-planes. These orientations can be arbitrary. Without loss of generality, we choose $S_i V_{j\pm} = \pm \delta_{ij} V_{j\mp}$. We can rotate the whole system to a canonical coordinate system $\tilde{x}$:

$$\begin{array}{ccc} \hat{K} = [-i\partial_1 \,\, -i\partial_2 \,\, -i\partial_4 \,\, -i\partial_5]^T & \& & \hat{X} = [x_1 \,\, x_2 \,\, x_4 \,\, x_5]^T \\ \tilde{K} = V_0 \hat{K} & \& & \tilde{X} = V_0^T \hat{X} \end{array}. \quad (S20)$$

The summation $\Sigma^2$ can be directly solved in this selected coordinate system:

$$\begin{aligned}
\Sigma^2 &= v_3^2 k_3^2 + v_\parallel^2 \cdot \left(\hat{K}^T \cdot \hat{K} + \hat{X}^T \hat{S}^T \hat{S} \hat{X} + 2\hat{X}^T \hat{S}^T \cdot \hat{K}\right) \\
&= v_3^2 k_3^2 + v_\parallel^2 \cdot \left(\tilde{K}^T \cdot V_0^T V_0 \cdot \tilde{K} - \tilde{X}^T V_0^T \hat{S}\hat{S} V_0 \tilde{X} - 2\tilde{X}^T \cdot V_0^T \hat{S} V_0 \cdot \tilde{K}\right) \\
&= v_3^2 k_3^2 + v_\parallel^2 \cdot \left[\tilde{K}^T \tilde{K} - \tilde{X}^T V_0^T (\theta_1^2 S_1^2 + \theta_2^2 S_2^2) V_0 \tilde{X} - 2\tilde{X}^T \cdot V_0^T (\theta_1 S_1 + \theta_2 S_2) V_0 \cdot \tilde{K}\right] \\
&= v_3^2 k_3^2 + v_\parallel^2 \cdot \left\{\tilde{K}^T \tilde{K} + \tilde{X}^T \begin{bmatrix} \theta_1^2 \cdot V_{1\pm}^T P_1 V_{1\pm} & 0 \\ 0 & \theta_2^2 \cdot V_{2\pm}^\dagger P_2 V_{2\pm} \end{bmatrix} \tilde{X} - 2\tilde{X}^T \cdot \begin{bmatrix} \theta_1 \cdot V_{1\pm}^T S_1 V_{1\pm} & 0 \\ 0 & \theta_2 \cdot V_{2\pm}^\dagger S_2 V_{2\pm} \end{bmatrix} \cdot \tilde{K}\right\} \\
&= v_3^2 k_3^2 + v_\parallel^2 \cdot \left\{\tilde{K}^T \tilde{K} + \tilde{X}^T \begin{bmatrix} \theta_1^2 \cdot I & 0 \\ 0 & \theta_2^2 \cdot I \end{bmatrix} \tilde{X} + 2\tilde{X}^T \cdot \begin{bmatrix} \theta_1 \cdot i\sigma_y & 0 \\ 0 & \theta_2 \cdot i\sigma_y \end{bmatrix} \cdot \tilde{K}\right\}
\end{aligned}$$

(S21)

The summation $\Sigma^2$ can be directly decomposed into two orthogonal models in Eq. S4, with $|B_{Pi}| = 2\theta_i$. This summation $\Sigma^2$ contains two sets of zero-point energy and landau levels from two lower dimensional WP counterparts, and it can be represented as:

$$\begin{aligned}
\Sigma^2 &= v_3^2 k_3^2 + v_\parallel^2 \cdot \left\{\begin{array}{l} +\left[-\tilde{\partial}_1^2 - \tilde{\partial}_2^2 + \theta_1^2 \cdot (\tilde{x}_1^2 + \tilde{x}_2^2) + 2\theta_1 \cdot \left(i\tilde{\partial}_2 x_1 - i\tilde{\partial}_1 x_2\right)\right] \\ +\left[-\tilde{\partial}_4^2 - \tilde{\partial}_5^2 + \theta_2^2 \cdot (\tilde{x}_4^2 + \tilde{x}_5^2) + 2\theta_2 \cdot \left(i\tilde{\partial}_5 x_4 - i\tilde{\partial}_4 x_5\right)\right] \end{array}\right\} \\
&= v_3^2 k_3^2 + v_\parallel^2 \cdot \left[2\theta_1 \cdot (2n_1 + 1) + 2\theta_2 \cdot (2n_2 + 1)\right] \\
&= v_3^2 k_3^2 + v_\parallel^2 \cdot \left[|B_{P1}| \cdot (2n_1 + 1) + |B_{P2}| \cdot (2n_2 + 1)\right]
\end{aligned}$$
(S22)

$n_1, n_2$ are two non-negative integers from the two sets of quantum harmonic oscillator models, and $|B_{Pi}| = 2\theta_i$. Furthermore, it can be verified that the generalized Zeeman effect can also be represented through these two effective magnetic fields:

$$\begin{aligned}
\pm \sqrt{\Xi + c_2 T_3 \cdot \text{sgn}(v_3) \Gamma_{3,ii}} &= 2v_\parallel^2 \cdot \text{diag}(-\theta_1 - \theta_2, +\theta_1 + \theta_2, +\theta_1 - \theta_2, -\theta_1 + \theta_2) \\
&= -v_\parallel^2 \cdot (|B_{P1}|\sigma_3 + |B_{P2}|\sigma_0) \otimes \tau_3
\end{aligned}$$
(S23)

Therefore, the YM under a gauge field contains similar supersymmetric Landau levels in the non-relativistic squared Hamiltonian with much richer degenerate features, as shown in Fig. 1(e):

$$\begin{aligned}
H_{YM,G}^2 &= v_3^2 k_3^2 + v_\parallel^2 \cdot \left[\begin{array}{l} 2\theta_1 \cdot (2n_1 + 1) + 2\theta_2 \cdot (2n_2 + 1) \\ +2 \cdot \text{diag}(-\theta_1 - \theta_2, +\theta_1 + \theta_2, +\theta_1 - \theta_2, -\theta_1 + \theta_2)\end{array}\right] \\
&= v_3^2 k_3^2 + v_\parallel^2 \cdot \left[|B_{P1}| \cdot (2n_1 + 1) + |B_{P2}| \cdot (2n_2 + 1) - (|B_{P1}|\sigma_3 + |B_{P2}|\sigma_0) \otimes \tau_3\right]
\end{aligned}$$
(S24)

The generalized Zeeman term contains two sets of Pauli matrices, where $\sigma$ and $\tau$ operate in inter-band $|i\rangle$ and intra-band $|\pm\rangle$ space, respectively. The generalized zero mode corresponds to the condition that both Zeeman lifts jointly cancel the zero-point energy induced by two sets of effective quantum harmonic oscillators, i.e., $-|B_{P1}|-|B_{P2}|=-\sqrt{\Xi+|T_3|}$, which means:

$$c_2 T_3 \cdot \text{sgn}(v_3)\Gamma_{3,ii} = |T_3| \Rightarrow \text{sgn}(v_3 c_2 T_3 \cdot \Gamma_{3,ii}) = 1. \tag{S25}$$

Here, similar to its low-dimensional counterpart, $\Gamma_{3,ii}$ is the diagonal element of the Gamma matrix $\Gamma_3$, and the condition $\Gamma_{3,ii} = 1$ or $\Gamma_{3,ii} = -1$ in Eq. S25 specifies that the eigenstate of the CZM is $|1,\pm\rangle$ or $|2,\pm\rangle$, respectively. This zero mode, shown by the red line in Fig. 1(f), has chiral group velocity determined by both the direction of the 4-form pseudovector field $T_3$ and the second Chern number $c_2$ of YM:

$$\omega_{CZM} = \text{sgn}(v_3 c_2 T_3 \cdot \Gamma_{3,ii}) \cdot \Gamma_{3,ii} v_3 k_3 = \text{sgn}(c_2 T_3) \cdot |v_3| \cdot k_3. \tag{S26}$$

However, it is worth noting that the condition $\text{sgn}(v_3 c_2 T_3 \cdot \Gamma_{3,ii}) = 1$ in Eq.S25 only guarantees the same Zeeman lift direction for both sets of quantum harmonic oscillators, thus it can only distinguish a special inter-band mode $|1\rangle$ or $|2\rangle$, but cannot specify the concrete intra-band state $|\pm\rangle$. The same Zeeman lift direction can jointly cancel the zero-point energy or conversely induce a larger band gap, i.e., $H^2_{YM,G} = v_3^2 k_3 + 2v_\parallel^2 \cdot [|B_{P1}|\cdot n_1 + |B_{P2}|\cdot n_2 + (|B_{P1}|+|B_{P2}|)]$ with $|B_{P1}|+|B_{P2}| = +\sqrt{\Xi+|T_3|}$. More notably, the same signs for $\Gamma_{3,ii}$ in intra-band $|\pm\rangle$ space imply that the eigenstate which induces a larger band gap can be degenerate with the zeroth eigenstate when the gauge field is removed. Therefore, the dispersion in Eq. S26 for CZM can only specify its group velocity, but cannot distinguish the concrete eigenstate, which is another unique feature of a higher dimensional YM. The eigenstates of these landau levels will be discussed later.

4. Particular cases of Landau levels in YM

This 4D rotation can be classified with respect to the values $\theta_1$ and $\theta_2$ as follows [1,2]:

a) If $\theta_1$ and $\theta_2$ are nonzero and $\theta_1 \neq \theta_2$, then the rotation generates double rotations;
b) If $\theta_1$ and $\theta_2$ are nonzero and $\theta_1 = \theta_2$, then the rotation generates isoclinic rotations;

c) If $\theta_2 = 0$ and $\theta_1 \neq 0$ or vice versa, then the rotation generates simple rotations;

We will mainly focus on the last two cases and discuss the corresponding Landau levels.

If $\theta_1 = \theta_2 = \theta \neq 0$ and $|B_{P1}| = |B_{P2}| = |B|$, it can be directly solved from Eq. S18 that the magnetic fields should satisfy the condition:

$$\Xi^2 = T_3^2 \Rightarrow \begin{cases} 1.\ B_{12} = +B_{45}, B_{15} = +B_{24}, B_{14} = -B_{25} \\ 2.\ B_{12} = -B_{45},\ B_{15} = -B_{24},\ B_{14} = +B_{25} \end{cases}. \quad \text{(S27)}$$

Such carefully chosen magnetic field components make the rotation planes no longer unique. There are infinitely many invariant planes instead of just two, called isoclinic rotations or Clifford displacements. This condition also induces a much higher degeneracy in Landau levels:

$$H_{YM,G}^2 = v_3^2 k_3^2 + v_\parallel^2 |B| \cdot [2 \cdot (n_1 + n_2) + \text{diag}(0, 4, 2, 2)]. \quad \text{(S28)}$$

YM has a more complicated Zeeman term. This result is similar to Eq. S6 with $H_{WP,G}^2 = v_3^2 k_3^2 + v_\parallel^2 |B_3| \cdot [2n + \text{diag}(0, 2)]$, but the integer number $n$ is replaced by $n_1 + n_2$.

This isoclinic rotation increases the degeneracy of Landau levels. In a non-relativistic squared Hamiltonian for a WP under a gauge field, the degeneracy in supersymmetric Landau levels only arises from the equality between the integer-spaced Landau level and the Zeeman splitting energy, which is protected by the first Chern number $c_1$. In a YM case, except for a more complicated Zeeman term, the degeneracy can also result from the SO(4) and SO(5) rotation symmetry, in which zero-point energy from different subspaces can be identical. For example, $n = 1$ in a WP system will correspond to two choices in a YM counterpart: $[n_1, n_2] = [0,1]$ and $[n_1, n_2] = [1,0]$. In fact, for any $n = N$ in a WP, there is always $N + 1$ degeneracy in a YM counterpart. Combining the above two degeneracy conditions, the degeneracy $g$ of the non-relativistic squared Hamiltonian at different Landau levels can be:

$$\omega^2 = \begin{cases} v_3^2 k_3^2, & g=1 \\ v_3^2 k_3^2 + 2v_\parallel^2 |B| \cdot N, & g=4N \end{cases}, \quad \text{(S29)}$$

as shown in Fig. 1(e), with $|B_{P1,P2}| = |B| \pm \Delta B$ ($\Delta B \ll B$) to distinguish the degeneracy.

Another special case satisfies $\theta_2 = 0$ and $\theta_1 \neq 0$ or vice versa, called simple rotation. This case can also be directly solved from Eq. S18:

$$T_3 = 0 \ \& \ \Xi \neq 0. \tag{S30}$$

In this case, the field component $B_{P2}$ is zero, and the system only has a nonzero magnetic field component $B_{P1}$. This condition induces Landau levels:

$$H_{YM,G}^2 = v_3^2 k_3^2 + v_\parallel^2 |B_{P1}| \cdot [2n_1 + \text{diag}(0, 2, 2, 0)]. \tag{S31}$$

It is also similar to Eq. S6. However, this system has two CZMs with opposite group velocities because the two zeros have opposite signs for $\Gamma_{3,ii}$ in Eq. S31. Besides, due to $\theta_2 = 0$, two dimensions that construct $P_2$ can be totally redundant. The 5D system is entirely the same as its lower dimensional counterpart -- a 3D 4-fold Dirac point. This 3D Dirac point is equivalent to overlapping two WPs with opposite topological charges $\pm c_1$ in the momentum space. Thus, the uniform magnetic fields will induce two opposite CZMs in different subspaces due to the opposite first Chern number $c_1$. That is indeed the helical zero modes in a Dirac system, protected by the first-topology $c_1$ in each sub-space instead of the nontrivial second-topology $c_2$ and induced by a 2-form magnetic field $\vec{B}$ instead of a 4-form pseudovector field $\vec{T}$.

It is worth mentioning that in both WP and YM systems under a gauge field, we have only solved the non-relativistic quadratic eigenequation $H_G^2 \Phi = \omega^2 \Phi$. The eigenvalues of a relativistic model can be obtained from these non-relativistic solutions based on chiral symmetry and the continuous condition. At $k_3 = 0$, both Hamiltonians satisfy the chiral symmetry due to the similar Clifford algebra:

$$\{H_{WP,G}, \sigma_3\} = \{H_{YM,G}, \Gamma_3\} = 0 \text{ at } k_3 = 0. \tag{S32}$$

This chiral symmetry ensures that for any state with energy $\omega_c > 0$, there is a chiral symmetric partner with energy $-\omega_c < 0$. Therefore, at $k_3 = 0$, except for the zeroth mode with $\omega_0 = 0$, the relativistic model has the same number of positive and negative eigenvalues, with degeneracy $g = 1$ for WP at $\omega = \pm v_\parallel \cdot \sqrt{N \cdot 2|B|}$ and degeneracy $g = 2N$ for YM at $\omega = \pm v_\parallel \cdot \sqrt{N \cdot 2|B|}$ with $|B_{P1}| = |B_{P2}| = |B|$. Away from $k_3 = 0$, the

dispersion should be continuous. Thus, for any eigenstates except for the zeroth mode, there are the same numbers of positive and negative eigenvalues located at $\pm\omega_c$, as shown in Fig. 1(d, f).

## 5. Eigenstates of Landau Levels in YM systems

In section 1, we have mainly focused on the dispersion spectra of WP and YM under a gauge field based on the non-relativistic squared Hamiltonians. The eigenstates of such a squared Hamiltonian are a series of Fock bases, which are the eigenstates of the particle number operator $\hat{n} = \hat{a}^\dagger \hat{a}$. However, the actual eigenstates of a relativistic semimetal under a gauge field can be linear combinations of these bases. The combination coefficients cannot be directly solved through the square Hamiltonians, except for the zeroth mode without any degeneracy.

In this section, inspired by the method of the fuzzy sphere [3-6], we introduce an exact mapping method to solve the eigenstates. Without loss of generality, here we only discuss a particular gauge setup with nonzero $A_4 = B_{24} \cdot x_2$ and $A_5 = B_{15} \cdot x_1$, which is indeed the experimental setup of the inhomogeneous Yang metamaterial. In the previous section, it has been demonstrated that any magnetic field setup in a YM can be turned into such a situation, with a specially chosen coordinate system. This 5D system only has two nonzero magnetic fields operating on two orthogonal 2-planes. Considering the experimental configuration, we also introduce an additional $k_3$-dependent onsite term along the axial direction, $\omega_{k0} = v_3(k_3 \mp K_{YM}) + \omega_p$, with $[\pm K_{YM}, \omega_p]$ the location of the Yang monopoles. Based on the derivation in the previous section, it is easy to prove that such a revision only tilts the dispersion spectra but does not change the eigenstates.

In this unique setup with a specially chosen coordinate system, the relativistic Hamiltonian in Eq. S9 can be written as:

$$H_{YM,G\pm} = \omega_{k0} - i \sum_{j=1}^{5} v_j \mathcal{D}_j \cdot \Gamma_j$$

$$= \omega_p I + v_3(k_3 \mp K_{YM}) \cdot (\Gamma_3 + I) + v_\parallel \cdot \begin{bmatrix} -i\partial_1 \cdot \Gamma_1 - i\partial_2 \cdot \Gamma_2 \\ +(k_4 - B_{24}x_2) \cdot \Gamma_4 + (k_5 - B_{15}x_1) \cdot \Gamma_5 \end{bmatrix}$$

(S33)

The five Hermitian Gamma matrices satisfying $\{\Gamma_i, \Gamma_j\} = 2\delta_{ij}$ equal to $\vec{\Gamma} = [-\sigma_1\tau_0, \sigma_2\tau_2, \sigma_3\tau_0, \sigma_2\tau_1, \sigma_2\tau_3]$ under the selected four polarization basis states

$\{|E_x\rangle, |E_y\rangle, |E_z\rangle, |H_z\rangle\}$, where $\sigma$ operates in the inter-band space, i.e., transverse and longitudinal mode, and $\tau$ operates in the intra-band space, i.e., $\{|E_x\rangle, |E_y\rangle\}$ and $\{|E_z\rangle, |H_z\rangle\}$. For brevity, the constant onsite energy $\omega_p$ will be ignored in subsequent calculations. Besides, it should be highlighted that in the 5D Landau level expression, the two partial derivative operators $k_4$ and $k_5$ will be redundant. Such redundant wavevectors are typical features of a magnetic response, which also exists in its low-dimensional counterpart. Thus, we can safely select $k_4 = k_5 = 0$ without loss of generality.

Here we define $B_{15} = s_p \alpha_p^2$ and $B_{24} = s_q \alpha_q^2$, with $s_p, s_q = \pm 1$ the sign and $\alpha_i = \sqrt{|B_i|} > 0$ the strength. Operators in two orthogonal 2-planes are denoted by index '$p$' and '$q$', respectively. Considering the fact that the algebraic structures in the in-plane degrees of freedom, e.g., the commutation relation $[x_1, -i\partial_1] = i$, is similar to a harmonic oscillator, one can replace both $x_1$ and $-i\partial_1$ by the ladder operators $\hat{a}_p^\dagger$ and $\hat{a}_p$: $x_1 = \frac{1}{\sqrt{2}\alpha_p} \cdot (\hat{a}_p + \hat{a}_p^\dagger)$ and $-i\partial_1 = \frac{\alpha_p}{\sqrt{2}i}(\hat{a}_p - \hat{a}_p^\dagger)$. Similarly, one can also replace $x_2$ and $-i\partial_2$ by $\hat{a}_q^\dagger$ and $\hat{a}_q$. These creation and annihilation operators satisfy the canonical *bosonic* commutation relation $[\hat{a}_p, \hat{a}_p^\dagger] = [\hat{a}_q, \hat{a}_q^\dagger] = 1$ and $[\hat{a}_p, \hat{a}_q^\dagger] = 0$. Therefore, Eq. S33 suggests a Hamiltonian tensor in the second quantization formulation to describe the inhomogeneous *classical* system:

$$\begin{aligned} H_{QM} &= v_3(k_3 \mp K_Y) \cdot (\Gamma_3 + I) + \frac{\sqrt{2}}{2} v_\parallel \cdot \begin{bmatrix} -\Gamma_1 \cdot \alpha_p (\hat{a}_p - \hat{a}_p^\dagger) i - \Gamma_5 \cdot \alpha_p s_p (\hat{a}_p + \hat{a}_p^\dagger) \\ -\Gamma_2 \cdot \alpha_q (\hat{a}_q - \hat{a}_q^\dagger) i - \Gamma_4 \cdot \alpha_q s_q (\hat{a}_q + \hat{a}_q^\dagger) \end{bmatrix} \\ &= U_0 + U_p \cdot \hat{a}_p + U_p^\dagger \cdot \hat{a}_p^\dagger + U_q \cdot \hat{a}_q + U_q^\dagger \cdot \hat{a}_q^\dagger \end{aligned}$$

(S34)

With the definitions $U_p = -v_\parallel \alpha_p (i\Gamma_1 + s_p \Gamma_5)/\sqrt{2}$, $U_q = -v_\parallel \alpha_q (i\Gamma_2 + s_q \Gamma_4)/\sqrt{2}$, and $U_0 = v_3(k_3 \mp K_{YM}) \cdot (\Gamma_3 + I)$. Such a Hamiltonian operates on the Hilbert space spanned by the orthonormal Fock states $|p, q\rangle = |p\rangle \otimes |q\rangle$, with $p, q = 0, 1, 2 \dots$ representing the number of bosonic particles in two orthonormal subspaces. These Fock states are defined in the real space as the direct production of a series of Hermite polynomials of order $p$ and $q$, that is:

$$|p,q\rangle = N_p H_p(\alpha_p x_1) \cdot N_q H_q(\alpha_q x_2) \cdot \exp\left[ik_3 x_3 - \frac{1}{2}(\alpha_p^2 x_1^2 + \alpha_q^2 x_2^2)\right], \quad (S35)$$

with $H_n(\xi) = (-1)^n e^{\xi^2} \frac{d^n}{dx^n} e^{-\xi^2}$ representing the Hermite polynomial of order $n$, and

$N_n = \sqrt{\frac{\alpha}{\pi^{1/2} \cdot 2^n \cdot n!}}$ representing the normalization factor.

It is worth mentioning that the real space variables $\{x_1, x_2\}$ and momentum space variables $\{k_1, k_2\}$ can no longer be distinguished separately in Eq. S34 and their performance can be similar. Such indistinguishability is the mathematical basis for the fact that a second-order singularity with a divergent 4-form momentum space curvature $\vec{F} \wedge \vec{F}$ and 4-form momentum space tensor gauge field $G_{1234}$ will directly couple to a 4-form real space external field $\vec{B} \wedge \vec{B}$ [6].

To rigorously solve this Hamiltonian, one can map this quantized Hamiltonian onto a quasi-3D non-Abelian tight-binding model, with a well-defined $k_3$ along $z$-direction and a semi-infinite 2D lattice in the virtual 2D space denoted by '$p$' and '$q$', as shown in Fig. S2(a). Mathematically, this mapping can be:

$$\begin{aligned}\hat{H}_{QM} &= \sum_{p,q,p',q'} \langle p',q'|\hat{H}_{QM}|p,q\rangle \cdot |p',q'\rangle\langle p,q| \\ &= \sum_{p,q} \sum_{\delta_p, \delta_q = 0, \pm 1} \langle p+\delta_p, q+\delta_q|\hat{H}_{QM}|p,q\rangle \cdot |p+\delta_p, q+\delta_q\rangle\langle p,q|\end{aligned} \quad (S36)$$

Since the ladder operators only connect two states with $\Delta p|\Delta q = \pm 1$, except for $\langle p,q|\hat{H}_{QM}|p,q\rangle = U_0$, this lattice model only includes the nearest-neighbor coupling terms between the site $|p,q\rangle$ and $|p\pm 1,q\rangle$ ($|p,q\pm 1\rangle$) along with the virtual horizontal (longitudinal) direction, i.e., $\langle p,q|\hat{H}_{QM}|p+1,q\rangle = U_p \cdot \sqrt{\lambda p + 1}$ and $\langle p,q|\hat{H}_{QM}|p,q+1\rangle = U_q \cdot \sqrt{\lambda q + 1}$, as shown in Fig. S2(b, c). Here, an additional coefficient $\lambda$ is added manually for the follow-up discussion, and $\lambda = 1$ in the exact model. Besides, due to $\hat{a}_i|0\rangle = 0$ for the vacuum state with no particles, this 2D lattice model physically has a hard boundary at the virtual lattice site $p = 0$ or $q = 0$. With the initially chosen orthogonal basis, $U_0$ does not have any intra-coupling terms and only has diagonal terms

representing the onsite energy. $U_0 = 0$ when $k_3 = \pm K_{YM}$ at the original location of YM. This discrete inhomogeneous non-Abelian tight-binding model in the virtual quasi-3D space can fully describe the inhomogeneous continuum 5D YM metamaterial system. The eigenvector $c_{pq,i}$ from this quasi-3D non-Abelian tight-binding model implies a field distribution:

$$\Phi = \exp(ik_3 x_3) \cdot \exp(ik_4 x_4 + ik_5 x_5) \cdot \sum_{i=1}^{4} \sum_{p=0}^{p_{max}} \sum_{q=0}^{q_{max}} |p,q\rangle \otimes \Psi_{YM,i} \cdot c_{pq,i}, \quad (S37)$$

with $\Psi_{YM}$ the basis of the Yang Monopole. It is worth noting that those nonzero $k_3$, $k_4$ and $k_5$ wavevectors, if present, will only introduce a trivial phase to the eigenstate.

It is worth mentioning that the size of the lattice model is actually semi-infinite, so a numerical solution must be manipulated with an appropriate numerical truncation. The numerical truncation applies two hard boundaries at those sites $|p_{max}, q\rangle$ and $|p, q_{max}\rangle$. Thus, those solutions localized near the virtual border must be removed due to the physically semi-infinite lattice model. As shown in Fig. S3, the eigenstates shown in Fig. S3(b-d) must be removed, and only the eigenstate shown in Fig. S3(g, h) is retained. This mechanism selects the actual physical field distribution and dispersion relation corresponding to the relativistic Hamiltonian.

Two specific spectra corresponding to $B_{15} = \pm B_{24}$ near $k_3 = K_{YM}$ are shown in Fig. S4(a, b). Clearly, only one CZM remains near the original YM. The preserved mode depends on the hybrid sign $s = s_p \cdot s_q$, which just corresponds to the sign of the aforementioned 4-form pseudovector field $T_3$. Near the positive YM located at $k_3 = K_{YM}$ with $c_2 = 1$, the sign $s = sgn(T_3) = \pm 1$ just corresponds to a preserved mode along the dispersion of the initial degeneracy horizontal/longitudinal modes. We show the lowest few eigenmodes at $k_3 = K_{YM}$ in Fig. S4(a, b). The zeroth mode always corresponds to a left-lower corner mode $|0,0\rangle$ in the 2D TBM model with nonzero $|E_x\rangle[|E_z\rangle]$ component in Fig. S4(c)[S4(d)]. In the real space, this state exhibits a pure centrally localized Hermite-Gauss field distribution denoted in Eq. S35 and illustrated in Fig. S4(e)[S4(f)]. Such a result is consistent with the result of a squared Hamiltonian.

Besides this zeroth mode, the spectrum also contains a series of degeneracy eigenstates. At $k_3 = K_{YM}$, these eigenfrequencies are located at $\omega_p \pm \sqrt{2n} \cdot v_\parallel \alpha$, with $n$ a positive integer, which are consistent with Eq. S28. These modes are indeed the bulk modes of the inhomogeneous 2D lattice model, with nonzero components located at a series of coupled nodes. These eigenstates satisfy the chiral symmetry, as shown in Eq. S32. As an example, the eigenstate $|\Phi\rangle$ in Fig. S4(g)[S4(h)] located at $\omega_p + \sqrt{2n} \cdot v_\parallel \alpha$ has a chiral symmetric partner $\Gamma_3 \cdot |\Phi\rangle$ located at $\omega_p - \sqrt{2n} \cdot v_\parallel \alpha$, as shown in Fig. S4(k)[S4(l)]. The same chiral pairs are also manifested in the spatial distribution of the field, as shown in Fig. S4(i, j) and S4(m, n).

Through this exact mapping method, not only the dispersion spectra, but also the concrete eigenstates can be directly solved. The results are fully consistent with the spectra discussed in the squared Hamiltonian. Interestingly, it is also worth mentioning that the huge matrix in Eq. S36 is blocked diagonal, as shown in Fig. S5. In fact, for the relativistic eigenstate at $\omega_C$, only those Fock states corresponding to $\omega_C^2$ in the non-relativistic squared Hamiltonian matter, as shown in Fig. S4(g-n). This exact mapping method is indeed equivalent to substituting all Fock states with eigenvalues $\omega_C^2$ into a first-order equation, diagonalizing the Hamiltonian to find the coefficients of the Fock states corresponding to $\omega_C$ and $-\omega_C$. Such a block diagonal property is also the reason why the solutions can always converge through finite modes, even though the coupling coefficient of the system keeps increasing. Indeed, diagonalizing each block directly can be a more efficient way to solve the spectra and eigenstates. Nevertheless, discussing the lattice model can clarify the whole topological properties, which will be addressed in the next section.

## 6. Discussion on Second Chern Number and Nested Chern Number

As $\lambda \to 0$, the semi-infinite lattice model becomes a corner of a periodic lattice model. Three typical periodic lattice models are shown in Fig. S6(a-c) for a fixed $B_{24} < 0$ and various $B_{15} = \{B_{24}, -B_{24}, 0\}$, respectively. The first two unit cells with $B_{15} = \pm B_{24}$ manifest themself as 2D lattices with $\pi$-flux through per intercell plaquette, but with different lattice arrangements. Thus, the topological origin of the corner state $|0,0\rangle$ can be described by the bulk-boundary correspondence of the second-order Nested Wilson loop [7,8]. Meanwhile, this corner state is indeed the generalized CZM protected by the second Chern number $c_2$. Therefore, this generalized CZM bridges two types of topological invariants: the second Chern number $c_2$ and the Nested Chern number defined on the Wannier band. In this section, we will discuss these topological properties in detail.

The effective bulk Hamiltonian of the periodic lattice is obtained from Eq. S34, through performing the Fourier transform for the virtual dimensions. For brevity, we set the lattice constant of the virtual quasi-3D lattice to 1, and the Fermi velocities $|v_\parallel| = |v_3| = 1$, and define $p_3 = k_3 \mp K_{YM}$. This effective bulk Hamiltonian can be represented:

$$\hat{h}_k = p_3(\Gamma_3 + I) - \sqrt{2}\alpha_p \cdot (\sin k_p \cdot \Gamma_1 + \cos k_p \cdot s_p \Gamma_5) - \sqrt{2}\alpha_q \cdot (\sin k_q \cdot \Gamma_2 + \cos k_q \cdot s_q \Gamma_4),$$
(S38)

where $k_p$ and $k_q$ are the momenta of two virtual dimensions.

The eigenvalues of the bulk Hamiltonian are double degenerate, with $\omega = p_3 + [-1, -1, +1, +1] \cdot p_0$, and $p_0 = (2\alpha_p^2 + 2\alpha_q^2 + p_3^2)^{1/2}$. This Hamiltonian represents a 3D topological insulator, with $\alpha_p \cdot \alpha_q \neq 0$. The two degenerate eigenstates corresponding to the negative frequencies can be expressed as:

$$|\Psi\rangle = |\Psi_0\rangle \cdot U = \sqrt{\frac{p_0 - p_3}{2p_0}} \cdot \begin{bmatrix} \tau_0 \\ R \end{bmatrix} \cdot U$$

$$R = \frac{\sqrt{2}}{p_0 - p_3} \cdot [-\alpha_p \sin k_p \cdot \tau_0 + i(\alpha_q s_q \cos k_q \cdot \tau_1 + \alpha_q \sin k_q \cdot \tau_2 + \alpha_p s_p \cos k_p \cdot \tau_3)]$$
(S39)

Here, $\tau_i$ represents the Pauli matrix, and the matrix $U$ represents an arbitrary $SU(2)$ gauge field satisfying $U^\dagger \Psi_0^\dagger \cdot \Psi_0 U = U^\dagger U = 1$.

These eigenstates indicate an $SU(2)$ non-Abelian Berry connection $\vec{A}_i = i\langle\Psi|\partial_{k_i}|\Psi\rangle = U^\dagger \bar{A}_i U + iU^\dagger \partial_{k_i} U$, with $\bar{A}_i = i\langle\Psi_0|\partial_{k_i}|\Psi_0\rangle$:

$$\bar{A}_p = \frac{\alpha_p}{p_0\,(p_0-p_3)} \cdot \begin{bmatrix} -\alpha_p s_p & -\alpha_q s_q \cdot e^{-i\theta_k} \\ -\alpha_q s_q \cdot e^{i\theta_k} & \alpha_p s_p \end{bmatrix}$$

$$\bar{A}_q = \frac{\alpha_q}{p_0\,(p_0-p_3)} \cdot \begin{bmatrix} -\alpha_q s_q & \alpha_p s_p \cdot e^{-i\theta_k} \\ \alpha_p s_p \cdot e^{i\theta_k} & \alpha_q s_q \end{bmatrix}$$

$$\bar{A}_3 = \begin{bmatrix} 0 & 0 \\ 0 & 0 \end{bmatrix} \tag{S40}$$

Here, $\theta_k = s_p k_p + s_q k_q$. The traces of these non-Abelian Berry connections are all zero, indicating that all bulk polarization along $k_p, k_q$ and $k_3$ are zero: $\vec{p} = (0,0,0)$.

To calculate the nested Wilson loop, firstly, we fix the gauge field $U = I$ and directly solve the non-Abelian Wilson loop from a base point $\vec{k}_B = (k_{pB}, k_{qB}, p_{3B})$:

$$\bar{W}_i(\vec{k}_B) = \mathcal{P} \exp\left[i\int_{\vec{k}_B}^{\vec{k}_B + 2\pi \cdot \vec{e}_i} \bar{A}_i \cdot dk_i\right] \tag{S41}$$

The operator $\mathcal{P}$ represents a path-ordered integral. Clearly, $\bar{W}_3 = 0$. However, in general, the analytic integrals require some special tricks. The integral can be:

$$\bar{W}_i(\vec{k}_B) = U(\theta_{kB} + 2\pi) \cdot W_i(\theta_{kB}, p_{3B}, U) \cdot U^\dagger(\theta_{kB})$$

$$= U(\theta_{kB} + 2\pi) \cdot \mathcal{P} \exp\left[i\int_{\theta_{kB}}^{\theta_{kB}+2\pi s_i} s_i \vec{A}_i(\theta_k, p_{3B}) \cdot d\theta_k\right] \cdot U^\dagger(\theta_{kB})$$

with $s_i \vec{A}_i(\theta_k, p_{3B}) = U^\dagger \cdot s_i \bar{A}_i(\theta_k, p_{3B}) \cdot U + iU^\dagger \partial_{\theta_k} U$ \hfill (S42)

Here, $\theta_{kB} = s_p k_{pB} + s_q k_{qB}$. Because $U$ is an arbitrary $SU(2)$ matrix, we can select a special $U$ to ensure the integration function to be $\theta_k$-independent:

$$U_{\theta_k} = \exp(-i\theta_k/2 \cdot \tau_3). \tag{S43}$$

The non-Abelian Berry connections for this particular gauge field $U = U_{\theta_k}$ are $\theta_k$-independent:

$$\vec{A}_p = -\frac{\alpha_p \alpha_q}{p(p-p_3)} \cdot s_q \tau_1 + \left[\frac{1}{2} - \frac{\alpha_p^2}{p(p-p_3)}\right] \cdot s_p \tau_3 = A_{p0} \cdot (\cos\eta_p \cdot \tau_1 + \sin\eta_p \cdot \tau_3)$$
$$\vec{A}_q = +\frac{\alpha_p \alpha_q}{p(p-p_3)} \cdot s_p \tau_1 + \left[\frac{1}{2} - \frac{\alpha_q^2}{p(p-p_3)}\right] \cdot s_q \tau_3 = A_{q0} \cdot (\cos\eta_q \cdot \tau_1 + \sin\eta_q \cdot \tau_3)$$
(S44)

Therefore, the non-Abelian Wilson loop $\bar{W}_i$ can be analytically solved:

$$\bar{W}_i = \bar{W}_i(\theta_B, p_{3B}, U=I) = U(\theta_B + 2\pi) \cdot \exp\left(i \cdot 2\pi s_i \cdot s_i \vec{A}_i\right) \cdot U(\theta_B)^\dagger$$
$$= -\exp(-i\theta_B/2 \cdot \tau_3) \cdot \exp\left(i \cdot 2\pi \vec{A}_i\right) \cdot \exp(+i\theta_B/2 \cdot \tau_3)$$
(S45)

This Wilson loop $W_i$ diagonalizes as:

$$\bar{W}_i = \sum_\pm |v_{\vec{k}_B}^{v_i^\pm}\rangle e^{2\pi i \cdot v_i^\pm} \langle v_{\vec{k}_B}^{v_i^\pm}| \quad \xleftrightarrow{|\tilde{v}_{\vec{k}_B}^{v_i^\pm}\rangle = U_{\theta_B}^\dagger |v_{\vec{k}_B}^{v_i^\pm}\rangle} \quad \tilde{A}_i = A_{i0} \cdot |\tilde{v}_{\vec{k}_B}^{v_i^\pm}\rangle \tau_z \langle \tilde{v}_{\vec{k}_B}^{v_i^\pm}|$$

with $v_i^\pm = \text{mod}\left(\frac{1}{2} \pm A_{i0}, 1\right)$ and $|v_{\vec{k}_B}^{v_i^\pm}\rangle = \frac{\sqrt{2}}{2} \cdot \sqrt{1 \mp \sin\eta_i} \cdot \begin{bmatrix} \frac{\sin\eta_i \pm 1}{\cos\eta_i} \\ e^{i\theta_B} \end{bmatrix} \cdot e^{i\phi_G}$
(S46)

Here, $|w_{\vec{k}_B}^{v_i^\pm}\rangle = |\Psi_0\rangle \cdot |v_{\vec{k}_B}^{v_i^\pm}\rangle = |\Psi\rangle \cdot |\tilde{v}_{\vec{k}_B}^{v_i^\pm}\rangle$ both represent the same eigenstate of a particular Wannier sector. The Wannier center is proportional to the phase $2\pi v_i^\pm$, with $v_i^- = mod(v_i^+, 1)$.

The nested Wilson loop and the nested Chern number can both be evaluated based on this gauge-dependent Wannier eigenstate $|w_{\vec{k}_B}^{v_i^\pm}\rangle$. These Wannier states have another $U(1)$ gauge phase $\exp(i\phi_G)$. However, both two invariants are gauge-independent, and we can use any gauge field that is convenient to achieve the same results. Thus, we will always fix this $U(1)$ gauge phase $\exp(i\phi_G) = 1$ in the following calculations.

### 7. The Nested Wilson Loop

At $p_3 = 0$, $\hat{h}_k$ has reflection symmetry $\hat{m}_i \hat{h}_k \hat{m}_i^\dagger = \hat{h}_{M_i k}$ for two virtual dimensions, where $\hat{m}_p = \sigma_2 \tau_0$, $\hat{m}_q = \sigma_1 \tau_2$ and $\{\hat{m}_p, \hat{m}_q\} = 0$. These symmetries quantize the nested

Wilson loop. Here, the Wannier eigenstate $|w_{\vec{k}_B}^{v_i^\pm}\rangle$ indicates an $U(1)$ abelian Wannier Berry connection:

$$\tilde{A}_j^{v_i^\pm} = i\langle w_{\vec{k}_B}^{v_i^\pm}|\partial_{k_j}|w_{\vec{k}_B}^{v_i^\pm}\rangle \begin{cases} = \langle v_{\vec{k}_B}^{v_i^\pm}|\bar{A}_j|v_{\vec{k}_B}^{v_i^\pm}\rangle + i\langle v_{\vec{k}_B}^{v_i^\pm}|\partial_{k_j}|v_{\vec{k}_B}^{v_i^\pm}\rangle \\ = \langle \tilde{v}_{\vec{k}_B}^{v_i^\pm}|\vec{A}_j|\tilde{v}_{\vec{k}_B}^{v_i^\pm}\rangle + i\langle \tilde{v}_{\vec{k}_B}^{v_i^\pm}|\partial_{k_j}|\tilde{v}_{\vec{k}_B}^{v_i^\pm}\rangle = \langle \tilde{v}_{\vec{k}_B}^{v_i^\pm}|\vec{A}_j|\tilde{v}_{\vec{k}_B}^{v_i^\pm}\rangle \end{cases}. \quad (S47)$$

These Wannier Berry connections can be analytically solved directly when $\alpha_p \cdot \alpha_q \neq 0$:

$$\tilde{A}_p^{v_q^\pm} = \begin{bmatrix} \dfrac{[\alpha_p^2 p_0 - \alpha_q^2 p_3 - (\alpha_p^2 + \alpha_q^2) \cdot p_{3p}] \cdot (p_0 + p_{3p})}{2\alpha_q^2 \cdot p_{3p} \cdot (p_3 + p_0)} \\ \dfrac{[-\alpha_p^2 p_0 + \alpha_q^2 p_3 - (\alpha_p^2 + \alpha_q^2) \cdot p_{3p}] \cdot (p_0 - p_{3p})}{2\alpha_q^2 \cdot p_{3p} \cdot (p_3 + p_0)} \end{bmatrix} \cdot s_p s_q, \quad (S48.1)$$

$$\tilde{A}_q^{v_p^\pm} = \begin{bmatrix} \dfrac{[-\alpha_p^2 p_3 + \alpha_q^2 p_0 - (\alpha_p^2 + \alpha_q^2) \cdot p_{3q}] \cdot (p_0 + p_{3q})}{2\alpha_p^2 \cdot p_{3q} \cdot (p_3 + p_0)} \\ \dfrac{[\alpha_p^2 p_3 - \alpha_q^2 p_0 - (\alpha_p^2 + \alpha_q^2) \cdot p_{3q}] \cdot (p_0 - p_{3q})}{2\alpha_p^2 \cdot p_{3q} \cdot (p_3 + p_0)} \end{bmatrix} \cdot s_p s_q, \quad (S48.2)$$

with $p_{3i} = \sqrt{2\alpha_i^2 + p_3^2}$. The Wannier Berry connection $\tilde{A}_j^{v_i^\pm}$ is only the function of $p_3$, $\alpha_p$, $\alpha_q$ and the hybrid sign $s = s_p \cdot s_q$, but independent of $k_p$ and $k_q$.

At $p_3 = 0$, these Wannier Berry connections are simply $\tilde{A}_p^{v_q^\pm} = \tilde{A}_q^{v_p^\pm} = -\dfrac{1}{2} \cdot s$ for arbitrary nonzero $\alpha_p$ and $\alpha_q$. Therefore, the Nested Wilson loop can be solved and quantized:

$$p_j^{v_i^\pm} = \dfrac{1}{(2\pi)^2} \int_{BZ} d^2k \cdot \tilde{A}_j^{v_i^\pm}. \quad (S49)$$

At $p_3 = 0$, the nested Wilson loop represents a well-defined Wannier polarization $\vec{p}^{v^\pm} = (p_p^{v_q^\pm}, p_q^{v_p^\pm}) = (1/2, 1/2)$ for each independent Wannier-sector and a quantized quadrupole moment $\vec{q}_{pq}^\pm = 2p_p^{v_q^\pm} \cdot p_q^{v_p^\pm} = 1/2$ [7,8]. Therefore, a corner-localized charge of 1/2 exists when the boundaries along both two virtual dimensions are open, that is the zero-mode localized at the natural corner site $|0,0\rangle$. The eigenstate of this corner mode depends on the concrete intercell coupling distribution. Due to the fact that $|0,0\rangle$ corresponds to the left-lower corner, the zeroth eigenstates are the $|E_x\rangle$ and $|E_z\rangle$ mode in Fig. S6(a) and S6(b), respectively, consistent with the numerical result shown in Fig. S4(a) and Fig. S4(b). Besides, it is worth noting that these reflection symmetries keep and quantize the quadrupole moment even when the lattice is anisotropic, i.e., $\alpha_p \neq \alpha_q$. Therefore, although a planar lattice structure cannot directly represent the non-Abelian internal coupling, and the bulk states are no longer 2-fold degenerate; the zeroth mode still exists with the eigenstate being the same as the isotropic one, as shown in Fig. S6(d) for various $\alpha_p$, with a tiny $U_0$ to distinguish the concrete zeroth mode.

In the special case with $\alpha_p \cdot \alpha_q = 0$, one or two Wannier polarizations tend to zero, $\vec{q}_{pq}^{\pm} = 2 p_p^{v_q^{\pm}} \cdot p_q^{v_p^{\pm}} = 0$. Therefore, the zeroth modes at $p_3 = 0$ manifest themselves as the edge states of a fully dimerized 1D SSH model instead of a corner state of a 2D lattice model, as shown in Fig. S6(c). As a result, two helical zero modes are presented near the original YM, which maintain the initial dispersions of both degeneracy modes, as shown in Fig. S6(f-h), which is consistent with Eq. S31.

## 8. The Nested Chern Number

Away from the original YM, i.e., $p_3 \neq 0$, the reflection symmetries are broken, and the quadrupole moment is no longer quantized, which is consistent with the result in Eq. S48. Therefore, changing the axial momentum variable, i.e., $p_3 \in (-\infty, \infty)$, is equivalent to the adiabatic charge pumping process arising exclusively from the bulk quadrupole moment. The varying quadrupole moment during this adiabatic charge pumping process can be mapped onto the chiral corner-localized hinge modes of a 3D semi-infinite insulator, which are topologically protected by two orthogonal but opposite nested first Chern numbers.

Through the above Wannier eigenstate $|w_{\vec{k}_B}^{v_i^{\pm}}\rangle$, we can also obtain the Wannier Berry connection along the axial direction: $\tilde{A}_3^{v_p^{\pm}} = \tilde{A}_3^{v_q^{\pm}} = 0$. Besides, these Wannier Berry connections $\tilde{A}_j^{v_i^{\pm}}$ are independent of $k_p$ and $k_q$. The general distributions of the Wannier Berry connection and Wannier Berry curvature along $p_3$ are shown in Fig. S7.

Thus, the nested Chern number defined on the Wannier sector $v_i^{\pm}$ can be obtained:

$$\begin{aligned}
c_{zp}^{v_{q,\pm}} &= \frac{1}{2\pi}\int_0^{2\pi} dk_p \int_{-\infty}^{\infty} dp_3 \cdot \tilde{F}_{zp}^{v_{q,\pm}} = \int_{-\infty}^{\infty} dp_3 \cdot \tilde{F}_{zp}^{v_{q,\pm}} \\
&= \int_{-\infty}^{\infty} dp_3 \cdot \left(\frac{\partial \tilde{A}_p^{v_{q,\pm}}}{\partial p_3}\right) = \tilde{A}_p^{v_{q,\pm}}\Big|_{p_3\to\infty} - \tilde{A}_p^{v_{q,\pm}}\Big|_{p_3\to-\infty} \\
&= \mp s
\end{aligned} \quad (\text{S}50.1)$$

$$\begin{aligned}
c_{qz}^{v_{p,\pm}} &= \frac{1}{2\pi}\int_0^{2\pi} dk_q \int_{-\infty}^{\infty} dp_3 \cdot \tilde{F}_{qz}^{v_{p,\pm}} = \int_{-\infty}^{\infty} dp_3 \cdot \tilde{F}_{qz}^{v_{p,\pm}} \\
&= \int_{-\infty}^{\infty} dp_3 \cdot \left(-\frac{\partial \tilde{A}_q^{v_{p,\pm}}}{\partial p_3}\right) = -\tilde{A}_q^{v_{p,\pm}}\Big|_{p_3\to\infty} + \tilde{A}_q^{v_{p,\pm}}\Big|_{p_3\to-\infty} \\
&= \pm s
\end{aligned} \quad (\text{S}50.2)$$

Therefore, $c_{zp}^{v_q^{\pm}} = -c_{qz}^{v_p^{\pm}} = \mp s = \mp s_p s_q$ for all nonzero $\alpha_p$ and $\alpha_q$. These two orthogonal but opposite nested first Chern numbers protect the chiral corner-localized hinge mode, which corresponds to the CZM, as shown in Fig. S6(e) [7,8]. Meanwhile, we have verified that this confined zero mode is protected by the second-order Chern number. Therefore, this mapping reveals the intrinsic connection between the nested first Chern number defined on the Wannier sectors $v_i^{\pm}$ and the original second Chern number defined on the 2-fold degeneracy conduction/valence bands.

In the special case with $\alpha_p \cdot \alpha_q = 0$, e.g., $\alpha_p = 0$ in the main text, the Wannier band $v_p^{\pm}$ is the same as the original band $|\Psi_0\rangle$, and consequently the nested first Chern number is the same as the one defined in each subspace. Meanwhile, the other Wannier band $v_q^{\pm}$ only has zero Berry connection. That is:

$$\tilde{A}_q^{v_p^{\pm}} = -\frac{p\pm p_3}{2p}\cdot s_q,\ \tilde{A}_p^{v_q^{+}} = 0 \Rightarrow c_{qz}^{v_{p,\pm}} = \pm s_q, c_{zp}^{v_{q,\pm}} = 0. \quad (\text{S}51)$$

Therefore, the zeroth modes are protected by the originally defined first-topology $c_{pz}$ in each sub-space instead of the nested first Chern number or the second Chern number. This result is also consistent with the discussion in section 1.3.

Finally, we will check the robustness of this topological protected CZM against defects through an additional nonuniformed magnetic field $B_{14}$ and $B_{25}$, i.e., $A_4 = B_{24}x_2 +$

$\Delta_A \cdot x_1^2$ and $A_5 = B_{15}x_1 + \Delta_A \cdot x_2^2$. In the quasi-3D lattice model, this setup corresponds to the next-nearest-neighbor coupling between the site $|p, q\rangle$ and $|p \pm 1, q \pm 1\rangle$ or $|p \pm 2, q\rangle$ ($|p, q \pm 2\rangle$) due to the nonzero $x_1^2$ and $x_2^2$ terms. However, the topologically protected CZM remains under limited perturbation, as shown in Fig. S8.

In summary, this generalized CZM is topologically protected by two equivalent topological properties: the second Chern number $c_2$ and the opposite nested first Chern numbers, $c_{zp}^{v_q^\pm}$ and $c_{qz}^{v_p^\pm}$ defined on the Wannier sectors.

## 9.  Discussion on Effective Medium Model and Simulation Model

Firstly, we discuss an ideal medium satisfying the constitutive equations:

$$\begin{bmatrix} \vec{D} \\ \vec{B} \end{bmatrix} = \begin{bmatrix} \varepsilon_0 \boldsymbol{\varepsilon} & (i\boldsymbol{\gamma}+\boldsymbol{\varsigma})/c \\ (-i\boldsymbol{\gamma}^\dagger+\boldsymbol{\varsigma}^\dagger)/c & \mu_0 \boldsymbol{\mu} \end{bmatrix} \begin{bmatrix} \vec{E} \\ \vec{H} \end{bmatrix}. \tag{S52}$$

This medium contains both bianisotropic $\boldsymbol{\gamma}$ and tellegen $\boldsymbol{\varsigma}$ terms:

$$\hat{\boldsymbol{\varepsilon}} = \begin{bmatrix} 1 & 0 & 0 \\ 0 & 1 & 0 \\ 0 & 0 & \varepsilon(\omega) \end{bmatrix}, \hat{\boldsymbol{\mu}} = \begin{bmatrix} 1 & 0 & 0 \\ 0 & 1 & 0 \\ 0 & 0 & \mu(\omega) \end{bmatrix}, \boldsymbol{\gamma} = \begin{bmatrix} 0 & 0 & \gamma_{xz} \\ 0 & 0 & \gamma_{yz} \\ -\gamma_{xz} & -\gamma_{yz} & 0 \end{bmatrix} \& \hat{\boldsymbol{\varsigma}} = \begin{bmatrix} 0 & \varsigma_{xy} & -\varsigma_{zx} \\ -\varsigma_{xy} & 0 & \varsigma_{yz} \\ \varsigma_{zx} & -\varsigma_{yz} & 0 \end{bmatrix}$$

(S53)

Besides, it satisfies a perfect electromagnetic duality with $\hat{\boldsymbol{\varepsilon}} = \hat{\boldsymbol{\mu}}$ and both parameters are described by the Drude model along z-direction: $\varepsilon(\omega) = \mu(\omega) = 1 - \omega_p/\omega^2$. Following the derivation in Ref. [9,10], a homogeneous medium can be mapped onto a 4-by-4 Hamiltonian to describe a 5D YM located at $\vec{k}_{YM} = [0,0,K_{YM},0,0]$ and $\omega = \omega_p$ with $K_{YM} = \omega_p/c$:

$$H = \frac{c}{2} \cdot (p_z - \omega_p \varsigma_{xy}) \cdot I + \frac{c}{2} \cdot \begin{bmatrix} (p_x - \omega_p \varsigma_{yz}) \cdot \Gamma_1 + (p_y - \omega_p \varsigma_{zx}) \cdot \Gamma_2 + (p_z - \omega_p \varsigma_{xy}) \cdot \Gamma_3 \\ + \omega_p \gamma_{xz} \cdot \Gamma_4 + \omega_p \gamma_{yz} \cdot \Gamma_5 \end{bmatrix}.$$

(S54)

Here, purely antisymmetric bianisotropic terms $\gamma_{xz} = -\gamma_{zx}$ and $\gamma_{yz} = -\gamma_{zy}$ serve as two synthetic wavevector dimensions $k_4$ and $k_5$, and purely antisymmetric tellegen terms $\varsigma_{ij} = -\varsigma_{ji}$ serve as shifts of three real wavevectors: $\Delta k_k = -\varepsilon_{ijk} \cdot \omega_p \varsigma_{ij}$.

It is worth mentioning that the synthetic momenta $\{k_4, k_5\}fo$ are continuous, unlike the Bloch momentum $\{k_x, k_y, k_z\}$ which are periodic. However, for the systems we are studying - Yang monopoles or five-dimensional gauge fields – it is only necessary for them to satisfy the dispersion relation within a small range. As a result, there is no distinction between the synthetic momenta and the real crystal momenta.

Furthermore, an inhomogeneous slow-varying medium can introduce an arbitrary gauge field $\vec{A}(\vec{r}) = A_i(\vec{r}) \cdot \hat{e}_i$ to the original YM based on a space-dependent shift of the YM locations, through gradient magneto-electric tensor including both antisymmetric bianisotropic and tellegen terms. Considering the fact that there are actually no coordinates of $x_4$ and $x_5$ in the synthetic model, here $\vec{r}$ can only represent the real space coordinates. Besides, these missing real-space dimensions also lead to the absence of two partial derivative operators along $x_4$ and $x_5$. Fortunately, as discussed in section 2 about Eq. S33, in a 5D Landau level system, two partial derivative operators will be naturally redundant. Thus, we can set $\partial_4 = \partial_5 = 0$ without loss of generality. The varying bianisotropic terms in synthetic dimensions actually represent the gauge potential $A_4 = -\omega_p \gamma_{xz}(\vec{r})$ and $A_5 = -\omega_p \gamma_{yz}(\vec{r})$ in an inhomogeneous metamaterial. Therefore, the redundancy exactly solves the problem associated with the absence of partial derivative operators, i.e., $-i\partial_4$ and $-i\partial_5$, in the synthetic model. Thus, the experimentally implemented inhomogeneous Yang metamaterial provides a general 5D YM with arbitrary *homogeneous* 2-form magnetic fields and 4-form pseudovector fields, while the measured single chiral mode is indeed the CZM in the overall 5-dimensional space, rather than just a projection/section of a subspace. Note that such a general configuration does not require any tellegen materials for observing the CZM because spatially varying bianisotropy terms $A_4 = B_{24} \cdot y$ and $A_5 = B_{15} \cdot x$ are sufficient to describe the most general cases in 5D. That's exactly the condition in the experiment.

Here, we will introduce a more detailed effective medium theory (EMT) with constitutive relation induced by the helical resonates to describe the bianisotropy in detail [9,10]. By considering the motion of electrons on each helical structure driven by external electromagnetic fields, the electromotive force can be written as:

$$\begin{cases} U = -i\omega L \cdot I + \dfrac{q}{C} + IR \\ U = \int d\vec{l} \cdot \vec{E} - \int d\vec{s} \cdot \partial \vec{B}/\partial t \end{cases} \& \begin{cases} I = \dot{q} = -i\omega q \\ \partial \vec{B}/\partial t = -i\omega \mu_0 \vec{H} \end{cases} \quad (S55)$$

where the RLC circuit model consists of a resistor $R$, an inductor $L$, and a capacitor $C$ to describe the effective response of the helical resonator. Besides, from the induced charge

$q$ and current $I$, one can write down other expressions for the electric polarizations and magnetic magnetization:

$$\vec{P} = q \cdot \vec{S}_p = q \cdot \int d\vec{l}, \quad \vec{M} = -i\omega q \cdot \vec{S}_m = I \cdot \int d\vec{s} \qquad (S56)$$

Here, we introduce two column vectors $\vec{S}_p$ and $\vec{S}_m$ to approximately represent the electric and magnetic dipole responses of a single helical resonator.

For a helical structure oriented in the $x$-direction, as shown in Fig. 2, these electric and magnetic dipole responses are represented as:

$$\vec{S}_p = [0, S_{p,y}, S_{p,z}]^T, \quad \vec{S}_m = [0, S_{m,y}, S_{m,z}]^T \qquad (S57)$$

In our design, the helix should be precisely adjusted to satisfy the constraints $S_{p,y}S_{m,z} + S_{p,z}S_{m,y} = 0$ and $|S_{p,z}| = K_{YM} \cdot |S_{m,z}|$, with optimized helical structure: $d = 1.0\ mm$, $r = 1.0\ mm$ and $\theta_{45} = 2.62\ rad$, as shown in Fig. 2(c). Besides, the size of one unit containing eight helices is $6.8mm \times 6.8mm \times 6.0mm$.

The electromotive force on the helix is: $U = (-\omega^2 L + 1/C - i\omega R)q = \vec{S}_p \cdot \vec{E} + i\omega\mu_0 \vec{S}_m \cdot \vec{H}$. With the effective RLC resonant frequency $\omega_0 = 1/\sqrt{LC}$ and effective loss $\Gamma = R/L$, the induced charge can be directly solved:

$$q = \frac{\vec{S}_p \cdot \vec{E} + i\omega\mu_0 \vec{S}_m \cdot \vec{H}}{L \cdot (\omega_0^2 - \omega^2 - i\omega\Gamma)}, \qquad (S58)$$

This charge induces both polarization $\vec{P}$ and magnetization $\vec{M}$ in Lorentz resonance formula:

$$\vec{P} = \frac{\vec{S}_p(\vec{S}_p \cdot \vec{E}) + i\omega\mu_0 \cdot \vec{S}_p(\vec{S}_m \cdot \vec{H})}{L \cdot (\omega_0^2 - \omega^2 - i\omega\Gamma)}, \quad \vec{M} = \frac{-i\omega\vec{S}_m(\vec{S}_p \cdot \vec{E}) + \omega^2\mu_0 \cdot \vec{S}_m(\vec{S}_m \cdot \vec{H})}{L \cdot (\omega_0^2 - \omega^2 - i\omega\Gamma)} \qquad (S59)$$

The unit cell of the Yang metamaterials is designed to contain a judicious combination of several resonant structures. Ignoring the coupling between different resonators, the total polarization field $\vec{P}$ and the magnetization field $\vec{M}$ are: $\vec{P} = (\sum_i V_i \vec{P}_i)/V$ and $\vec{M} = (\sum_i V_i \vec{M}_i)/V$, where $V_i$ and $V$ are the volume of the resonator and the volume of the metamaterial unit cell, respectively. Inserting this equation into the constitutive relation $\vec{D} = \varepsilon_0 \varepsilon_d \vec{E} + \vec{P}, \vec{B} = \mu_0(\vec{H} + \vec{M})$, one can derive its constitutive relation.

In the designed Yang metamaterial, four precisely adjusted helical structures combined with their mirror counterparts are collectively rotated to the angles $\Phi_{1 \mapsto 4} = \psi_{45} + [+\delta_{45}, +\delta_{45}+90°, -\delta_{45}+180°, -\delta_{45}+270°]$. Considering the transformation under a particular symmetry, a general dipole response for a rotated/mirrored helical unit can be:

$$\vec{S}_p(\phi_{Rot}, n_{Mir}) = M_z^{n_{Mir}} R_z(\phi_{Rot}) \cdot \vec{S}_p(0,0)$$
$$\vec{S}_m(\phi_{Rot}, n_{Mir}) = \det(M_z^{n_{Mir}}) \cdot M_z^{n_{Mir}} R_z(\phi_{Rot}) \cdot \vec{S}_m(0,0)$$
(S60)

Here, $R_z(\phi_{Rot}) = [\cos(\phi_{Rot}) - \sin(\phi_{Rot})\ 0; \sin(\phi_{Rot})\ \cos(\phi_{Rot})\ 0; 0\ 0\ 1]$ and $M_z = \text{diag}([1,1,-1])$ represent standard rotation and mirror operators, respectively. Therefore, the constitutive relation of Yang metamaterial can be expressed in the general formula:

$$\begin{bmatrix} \vec{D} \\ \vec{B} \end{bmatrix} = \begin{bmatrix} \varepsilon_0 \boldsymbol{\varepsilon} & i\boldsymbol{\gamma}/c \\ -i\boldsymbol{\gamma}^\dagger/c & \mu_0 \boldsymbol{\mu} \end{bmatrix} \begin{bmatrix} \vec{E} \\ \vec{H} \end{bmatrix},$$
(S61)

With:

$$\boldsymbol{\varepsilon} = \begin{bmatrix} \varepsilon_p & 0 & 0 \\ 0 & \varepsilon_p & 0 \\ 0 & 0 & \varepsilon_t \end{bmatrix}, \quad \begin{cases} \varepsilon_p = \varepsilon_d \cdot \left[1 + \frac{1}{2}\left(\frac{S_{m,y}}{S_{m,z}}\right)^2 \cdot \frac{\omega_p^2 - \omega_0^2}{\omega_0^2 - \omega^2 - i\omega\Gamma}\right] \\ \varepsilon_t = \varepsilon_d \cdot \left[1 + \frac{\omega_p^2 - \omega_0^2}{\omega_0^2 - \omega^2 - i\omega\Gamma}\right] \end{cases}$$
(S62.1)

$$\boldsymbol{\mu} = \begin{bmatrix} \mu_p & 0 & 0 \\ 0 & \mu_p & 0 \\ 0 & 0 & \mu_t \end{bmatrix}, \quad \begin{cases} \mu_p = 1 + \frac{1}{2}\left(\frac{S_{m,y}}{S_{m,z}}\right)^2 \cdot \frac{\omega^2 \cdot (\omega_p^2 - \omega_0^2)}{\omega_p^2 \cdot (\omega_0^2 - \omega^2 - i\omega\Gamma)} \\ \mu_t = 1 + \frac{\omega^2 \cdot (\omega_p^2 - \omega_0^2)}{\omega_p^2 \cdot (\omega_0^2 - \omega^2 - i\omega\Gamma)} \end{cases}$$
(S62.2)

$$\boldsymbol{\gamma} = \begin{bmatrix} 0 & 0 & \gamma_{xz} \\ 0 & 0 & \gamma_{yz} \\ -\gamma_{xz} & -\gamma_{yz} & 0 \end{bmatrix} \quad \& \quad \begin{cases} \omega_p \gamma_{xz} = -\frac{\sqrt{2\varepsilon_d}}{2} \cdot \frac{S_{m,y}}{S_{m,z}} \cdot \frac{\omega \cdot (\omega_p^2 - \omega_0^2)}{\omega_0^2 - \omega^2 - i\omega\Gamma} \cdot \sin\delta_{45} \cdot \cos(\psi_{45} + \pi/4) \\ \omega_p \gamma_{yz} = -\frac{\sqrt{2\varepsilon_d}}{2} \cdot \frac{S_{m,y}}{S_{m,z}} \cdot \frac{\omega \cdot (\omega_p^2 - \omega_0^2)}{\omega_0^2 - \omega^2 - i\omega\Gamma} \cdot \sin\delta_{45} \cdot \sin(\psi_{45} + \pi/4) \end{cases}$$

(S62.3)

Here, $\varepsilon_d = 1.8$ is the dielectric constant of the PCB board. In the model, $\omega_p = K_{YM} \cdot c/\sqrt{\varepsilon_d}$, $S_{p,z} = S_{m,z} \cdot K_{YM}$ and $S_{p,y} = -S_{m,y} \cdot K_{YM}$.

From this dispersive effective medium model, near the 4-fold degeneracy YM point, partially taking into account the effect of lattice scattering, we can derive the following 4-by-4 effective Hamiltonian using the $k \cdot p$ approximation method:

$$H_{YM} = \begin{bmatrix} 2v_3 p_3 & 0 & -v_\parallel (k_x + ik_5) & -v_\parallel (k_y + ik_4) \\ 0 & 2v_3 p_3 & -v_\parallel (-k_y + ik_4) & v_\parallel (-k_x + ik_5) \\ v_\parallel (-k_x + ik_5) & v_\parallel (k_y + ik_4) & 0 & 0 \\ v_\parallel (-k_y + ik_4) & -v_\parallel (k_x + ik_5) & 0 & 0 \end{bmatrix} \quad (S63)$$

with effective Fermi velocities: $v_3 = \dfrac{c}{2\sqrt{\varepsilon_d}}$ and $v_\parallel = v_3 \cdot \dfrac{\sqrt{\omega_p^2 - \omega_0^2}}{\sqrt{\omega_p^2 + \omega_0^2}}$, and effective synthetic wavevectors: $k_4 + ik_5 = -\dfrac{\sqrt{2}}{2} \dfrac{m_p}{m_z} \cdot K_{YM} \sin\delta_{45} \cdot \exp[i(\psi_{45} + 45°)]$. This Hamiltonian is just Eq. S33 in the preview discussion.

Clearly, the two rotation angles $[\delta_{45}, \psi_{45}]$ only vary the magneto-electric tensor $\boldsymbol{\gamma}$, but do not affect the relative permittivity and permeability tensors $\boldsymbol{\varepsilon}$ and $\boldsymbol{\mu}$. In a $k \cdot p$ approximation Hamiltonian, these purely antisymmetric bianisotropic terms can be viewed as two synthetic wavevectors or gauge fields, and the medium just behaves like a 5D YM with 4-fold degeneracy. By fitting suitable parameters, this constitutive relation in Eq. S62 can well describe the response of the Yang monopole, as shown in Fig. S9. The fitting parameters are close to the actual structural parameters:

$$\begin{aligned} \vec{S}_p &= \int d\vec{l} = \left[0, -2\bar{r}\sin\bar{\theta}_{45}, \bar{d}\right], \\ \vec{S}_m &= \int d\vec{s} = \left[0, -\bar{r}\bar{d} \cdot \left(1 - \frac{1}{2}\cos\bar{\theta}_{45}\right), \bar{r}^2 \cdot (2\pi - \bar{\theta}_{45})\right] \\ \bar{d} &= d = 1.0\,mm, r \approx \bar{r} = 0.925\,mm, \theta_{45} \approx \bar{\theta}_{45} = 3.45\,\text{rad} \\ \omega_p &= 2\pi \times 14.66\,GHz, \omega_0 = 2\pi \times 13.45\,GHz, \varepsilon_d = 1.8, \Gamma = 0.01\omega_p \\ P_x &= P_y = 6.8\,mm, P_z = 6.0\,mm \approx \bar{P}_z = 7.0\,mm \end{aligned} \quad (S64)$$

Here, the analytical solutions of the dipole models come from the rough approximation of a uniform current in the helical resonant.

Through an appropriate design of the spatial distribution of the two rotation angles $[\delta_{45}, \psi_{45}]$, any slow-varying gauge field can be realized in an inhomogeneous photonic metamaterial. The experimental sample possesses a linearly varying $\sin\delta_{45}$, which varies from 0 to 1 through 20 units, and a space-dependent phase distribution $\psi_{45} = atan2(x, y) - 45°$. The sample size is $275.4\ mm \times 275.4\ mm$ in the $x$-$y$ plane, with two 3 $mm$-wide vertical slits through the center of the sample, as shown in Fig. 3(a) and Fig. S10. Each unit along the $z$-direction contains 4 PCB layers, with two 1 $mm$-thick helical layers and two 2 $mm$-thick bare PCB layers. This designed inhomogeneous Yang metamaterial contains a uniform effective magnetic flux density $B_{15} = B_{24} \approx -1210\ m^{-2}$. These magnetic flux densities induce a uniform pseudovector field $T_3 \approx (0.011\omega_p)^4$ along the $z$-direction. A bandgap of 0.39GHz [$\Delta\omega = 2\sqrt{2}v_\parallel \alpha = 2\sqrt{2} \cdot (T_3/2)^{1/4} \approx 0.027\omega_p$] induced by such a strong pseudovector field can be clearly detected in the experiment.

The spectrum of this inhomogeneous metamaterial can be similarly solved through the exact mapping method. In the calculation, we will take fields of a fixed frequency $\omega$, and rewrite Maxwell's equations compactly in a 6-by-6 matrix formula $M\Phi = k_z \cdot N\Phi$, with:

$$M = \begin{bmatrix} \omega\varepsilon_0 \boldsymbol{\varepsilon} & i\omega\boldsymbol{\gamma}/c_0 - \hat{D}_{XY} \\ -i\omega\boldsymbol{\gamma}^+/c_0 + \hat{D}_{XY} & \omega\mu_0\boldsymbol{\mu} \end{bmatrix}, N = \begin{bmatrix} 0 & \lambda_z \\ -\lambda_z & 0 \end{bmatrix}, \Phi = \begin{bmatrix} \vec{E} \\ \vec{H} \end{bmatrix}$$

$$\hat{D}_{XY} = \begin{bmatrix} 0 & 0 & i\partial_y \\ 0 & 0 & -i\partial_x \\ -i\partial_y & i\partial_x & 0 \end{bmatrix} \quad \& \quad \lambda_z = \begin{bmatrix} 0 & 1 & 0 \\ -1 & 0 & 0 \\ 0 & 0 & 0 \end{bmatrix}$$

(S65)

Therefore, due to the spatial distribution of the bianisotropic terms $\boldsymbol{\gamma} = \boldsymbol{\gamma}(x, y)$, the operator $M$ can be written as a function of operators $-i\partial_x, -i\partial_y$ and $x, y$, and can be represented in the second quantization formulation, just like Eqs. S34-S37. This Hamiltonian can be similarly solved through the exact mapping method shown in Section 2. The results on the global dispersion and field distribution are shown in Fig. 2(d) and Fig.

2(e), which match well with the results obtained from the effective 4-by-4 Hamiltonian shown in Figs. S2-S8.

Finally, a FEM simulation for the polarization-dependent internal field is also performed based on the EMT model in Eq. S62 to simulate the experimental setup. The simulation is implemented using the weak form module of *Comsol* software, with a manually defined balanced weak form to solve the response of a finite-size medium with an inhomogeneous magneto-electric tensor:

$$\tilde{I}_{wf} = -\int_V d\tau \left\{ \begin{array}{c} \left(-k_0 \hat{\gamma}^\dagger \vec{G} + \nabla \times \vec{G}\right) \cdot \hat{\mu}^{-1} \cdot \left(-k_0 \hat{\gamma}^\dagger \vec{E} + \nabla \times \vec{E}\right) \\ -k_0^2 \vec{G} \cdot \hat{\varepsilon} \vec{E} \end{array} \right\} \quad (S66)$$

***G*** is the test function, by default, selected as the basis functions used to interpolate the electric field. The simulated results display a clear difference in excitation efficiency near $k_z = \pm K_{YM}$ and $\omega = \omega_p$ for $E_x$- /$E_y$- polarized excitations, which are consistent with the spectrum of the CZMs, as shown in Fig. 3 and Figs. S11-S14.

It is worth noting that we give priority to fitting the frequency in Fig. S9, which causes the thickness parameter to be slightly different from the real situation. Besides, when the resonant frequency $\omega_0$ is close to the plasmonic frequency $\omega_p$, $K_{YM}$ in the EMT model can also slightly shift from the $k \cdot p$ approximation result. These differences together induce the position of Yang monopole $K_{YM}$ to be slightly different in the simulation and experiment, and we will deal with this normalization parameter individually in Fig. S12. Fortunately, this difference does not affect the overall conclusion of the article. A similar irrelevant simulation degree of freedom term is the absorption of EMT. In fact, the overall response is the same regardless of the strength of the absorption. In the simulation, we choose $\Gamma = 0.01\omega_p$, which will induce a more pronounced contrast in Fig. S14, but it does not affect the overall conclusion of the article. Finer tuning of the simulation parameters could provide results much closer to the experimental values, but for cost and necessity reasons, the current results are kept.

## 10. Discussion on Experimental Measurements

As shown in Fig. 2(d), the generalized 5D CZM corresponds to a specific localized polarized mode in a 3D inhomogeneous Yang metamaterial. Thus, a polarization-dependent dispersion spectrum can be measured to verify this CZM.

The experimental setup is shown in Fig. S10. A polarization-dependent plane wave is launched by a horn antenna (HOAN 120G20A6S) located below the stacked metamaterial. The size of the antenna outlet face is about $80mm \times 100mm$, with a gain 20 dBi. This horn antenna ensures that the excitation in the x-y plane is concentrated around $k_x, k_y \approx 0$ and has a well-defined polarization. The polarized field distribution inside the sample is detected by the near-field scanning of a monopole antenna pointed to a specific direction: $x$, $y$, or $z$ direction. The exposed part of the monopole antenna is about 8 mm in length and about 0.5 mm in diameter, while the cladding part is about 14 cm in length and about 2.2 mm in diameter. Such a setup can provide a polarization selection that differs by more than ten times between co-polarization and cross-polarization.

For the characterization of the metamaterial, the field distribution inside the metamaterial is detected through two 3mm-width vertical slits cut through the center of the sample. We always use a co-polarization setting and adjust the polarization angle of both the horn and the monopole for measuring the field inside different slits. Specifically, considering the geometry of the vertical slits and the horizontal monopole antenna, we detect $E_x$-polarized field in the $x - z$ slit with $y = 0$, and $E_y$-polarized field in the $y - z$ slit with $x = 0$. Only half of the 160-layer samples (20 out of 40 units in total) we prepared contain the measurement slits. Therefore, the field distributions in Fig. 3 and Fig. S11 correspond to two individual measurements. The data $z \in (0,20)$ units come from the measurement for an 80-layer sample, while the data $z \in (20,40)$ units come from another measurement for a 160-layer sample. The results from the two measurements are consistent with each other and with the corresponding simulation results, in terms of propagation behaviors and wavevectors.

This inhomogeneous metamaterial only supports a propagating $E_x$-polarized CZM mode, but behaves as a bandgap for the $E_y$-polarized field, although the bianisotropic terms appear to be quasi-isotropic without a particular direction. Fig. S11 clearly shows this polarization dependent transmission feature. Notably, Fig. S11(d) shows a gradual increase in the decay length away from the plasmonic frequency at about 14.66 GHz. Far from the plasma frequency, the transmission features are similar for both polarizations, in both the frequencies around 14.1 GHz and 15.2 GHz with propagating waves and around the resonant frequency of 13.45 GHz with decaying waves. It is worth noting that the center where the field should be strongest is disturbed by the crossover slits. Besides, as shown in Fig. S11(d), the decay behavior of $E_y$ field is slightly asymmetrical in frequency, where the left side is slightly blue-shifted compared to the right side. This frequency asymmetry also results in a spatially asymmetric field distribution in fixed frequency measurements. This asymmetry is mainly due to the slight deviation of the parameters from the model under a wide range of designs. However, the overall contrast does not change with the slight imperfections of the sample.

A Fourier transformation is implemented to the simulated/measured field distributions to verify the polarization dependent dispersion, as shown in Fig. S12. The simulation results are shown in Fig. S12(a) and Fig. S12(d) for different polarizations. They show significant differences near the plasmonic frequency. The dispersion of $E_x$-polarized field is almost along the dispersion of the bulk, while the dispersion of $E_y$-polarized field exhibits a bandgap around the plasma frequency at about 14.66 GHz. Such a contrast is consistent across a series of calculations at different in-plane locations, except for the points far away from the center of the sample, e.g., the location of unit 15. This contrast is the direct result of the existence of a center-localized $E_x$-polarized CZM. The Fourier transformations are independently implemented in two individual measurements, as shown in Fig. S12(b) and S12(e) for $z \in (0,20)$ units and Fig. S12(c) and S12(f) for $z \in (20,40)$ units. The thicker configuration suffers more signal-to-noise ratio loss due to the stronger attenuation. Besides, the data at the crossover slits appear to be strongly disturbed. However, the contrast between the two polarizations occurs consistently across a series of measurements at different locations and for both sample configurations, with a gap for $E_y$-field near the plasmonic frequency at about 14.66 GHz. The slight imperfections of the sample do not change this overall contrast. Meanwhile, away from this plasmonic frequency, the dispersions for two polarizations are similar - both contain the bandgap near the resonance frequency at about 13.45 GHz, and the dispersions of the propagating modes are similar.

A similar contrast can also be detected through the transmitted signal, as shown in Fig. S13. Since this CZM is a bulk state, we can directly measure the transmitted field outside the sample. The $x-y$ plane measurement is implemented about 15 mm above the top surface for the 160-layer (40-Unit) sample, as shown in Fig. S13. Clearly, near the plasmonic frequency, only $E_x$-field can propagate, while the transmitted $E_y$-field is almost negligible. Meanwhile, the transmissions for the two polarizations look similar away from the plasmonic frequency.

Finally, we confirm the polarization dependent transmittance contrast through the measured transmitted power, as shown in Fig. S14. We calculate the transmitted power by integrating the line field in Fig. S11: $P_{Line} = 2\pi \cdot \int_{Line} |S_{21}|^2 \cdot r dr$, or integrating the surface field in Fig. S13: $P_{Surf} = \int_{Surf} |S_{21}|^2 \cdot d\vec{s}$, for different sample configurations. These two methods offer almost the same contrast in the simulation shown in Fig. S14(a) and Fig. S14(d). In a line-integral, the transmitted power can be normalized through two integrals at the top and the bottom surfaces, as shown in Fig. S14(b, c) for different sample configurations. Meanwhile, a surface integral only offers the absolute transmitted power, as shown in Fig. S14(e, f). Considering that the horn antenna used for different polarizations is the same, the total incident power is approximately the same here. In all four measurements, the transmitted powers of different polarizations always show noticeable differences near the plasmonic frequency, where the contrast can be about 25-30 dB in line-integral data and 20-30dB in surface-integral data.

In summary, by performing polarization resolved experimental measurements, combined with theory and simulation, we proved the existence of the polarization dependent localized mode in the inhomogeneous Yang metamaterial, which is direct evidence for the presence of 5D CZM arising from the 4-form pseudovector field.

# Figures

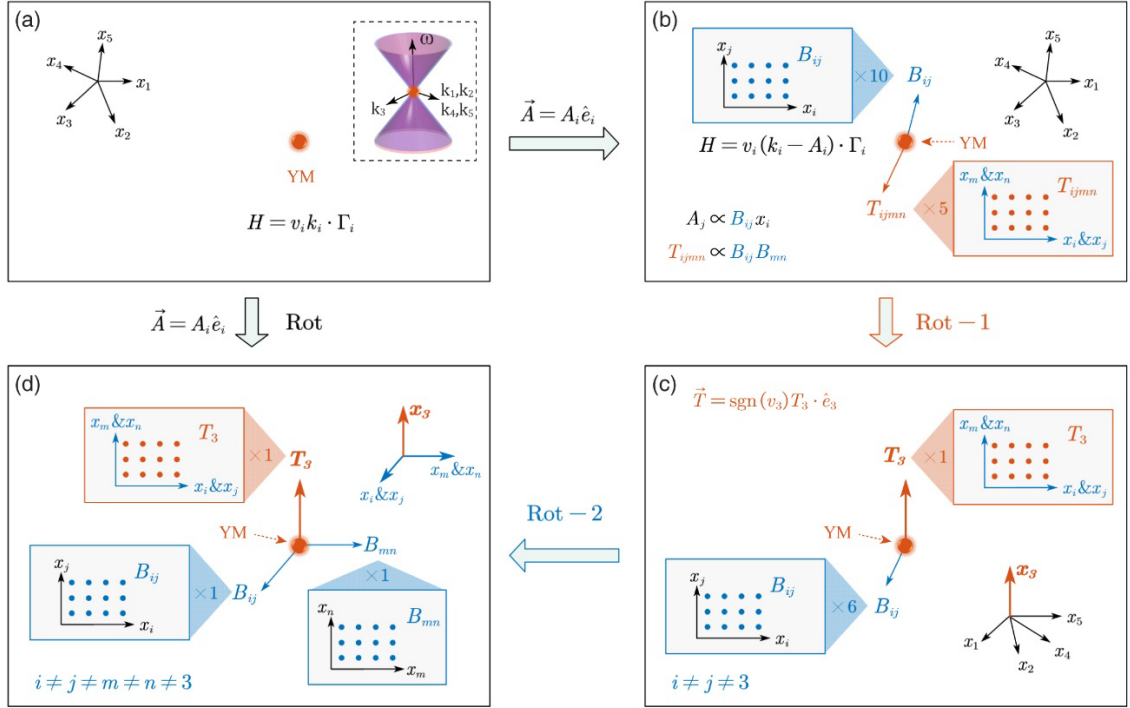

**Fig. S1. Schematic diagram of a YM under a gauge field .** (a) The original 5D YM with linear degenerate dispersion spectra. (b) In general, a YM can couple with ten 2-form magnetic field components $B_{ij}$ and five 4-form pseudovector field components $T_{ijmn} = T_k \propto \varepsilon^{ijkmn} B_{ij} B_{mn}$. (c) Without loss of generality, the direction of the pseudovector $\vec{T}$ can be chosen as the axial direction. (d) Orthogonal transformation for the 4D Euclidean space perpendicular to $x_3$ can separate it into a pair of 2-planes with orthogonal and individual magnetic fields: $B_{ij}$ and $B_{mn}$. Through a series of basis transformations, a YM only effectively experiences at most two perpendicular 2-form magnetic fields and one 4-form pseudovector field.

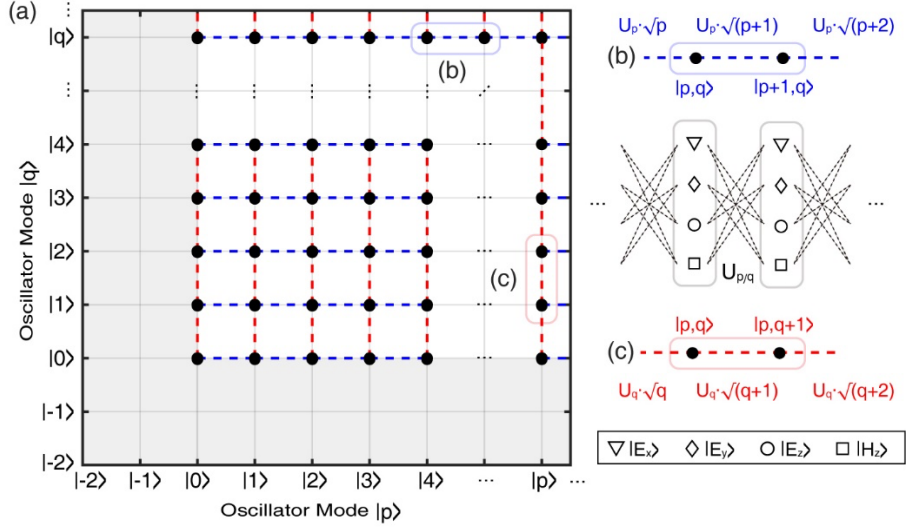

**Fig. S2. Schematic diagram of a quasi-3D lattice model, onto which an inhomogeneous 5D Yang metamaterial can be mapped.** (a) Illustration of the 2D semi-infinite non-Abelian lattice model, with hard boundaries at $p = 0$ and $q = 0$ enforced by $\hat{a}_i|0\rangle = 0$ for the vacuum state. (b, c) The general non-Abelian coupling configuration between the nearest-neighbor sites, along both the virtual (b) horizontal and (c) longitudinal direction.

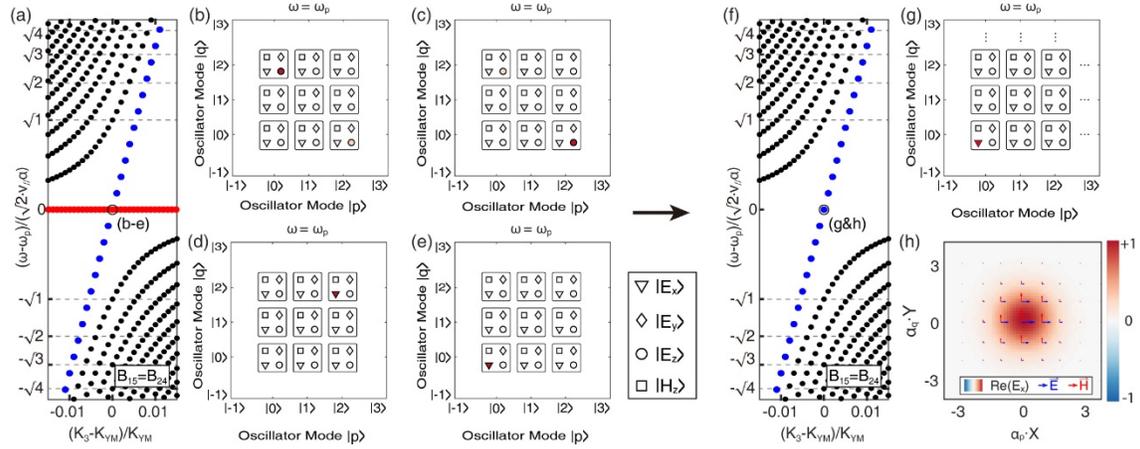

**Fig. S3. Illustration of eigenmodes with/without non-physical solutions.** (a) The dispersion of the quasi-3D lattice model, with $B_{15} = B_{24} = B < 0$. (b-e) The mode profiles in the virtual lattice for all four zeroth modes at $k_3 = K_{YM}$ and $\omega = \omega_p$, with $p_{max} = q_{max} = 2$ as an example. (f) The physical dispersion for the semi-infinite quasi-3D lattice model. (g, h) The mode distributions in the virtual lattice and in real space for the CZM.

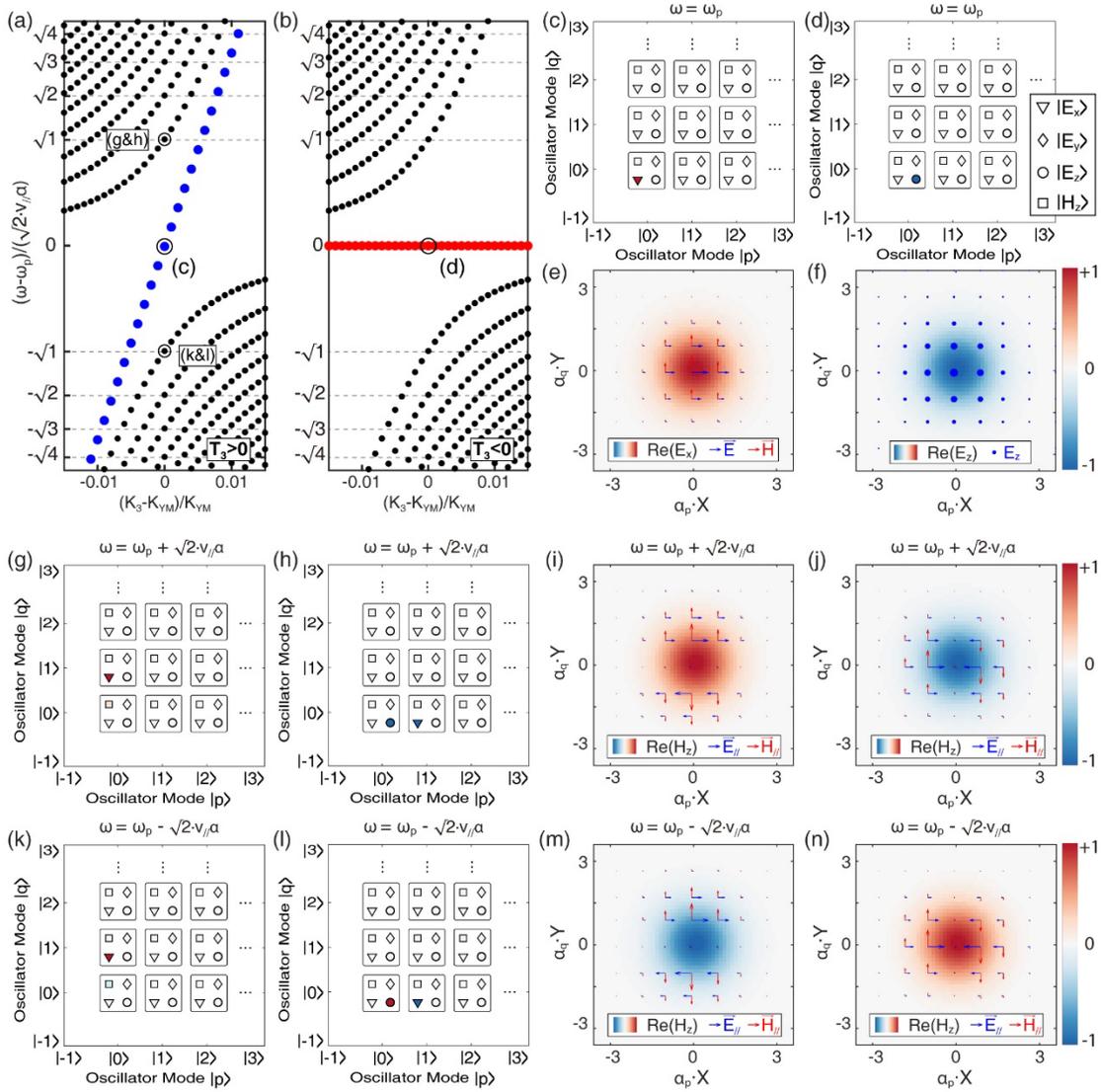

**Fig. S4. Dispersion and field distributions of typical eigenstates.** (a, b) The dispersion spectra for two typical setups with (a) $T_3 > 0$ and (b) $T_3 < 0$ and $|B_{15}| = |B_{24}|$. The blue/red dots represent the dispersion of the CZM. (c, d) The mode profiles in the virtual lattice for CZMs in (a, b). (e, f) The real-space field distributions correspond to (c, d). (g-n) Eigenstate profiles of typical bulk states labeled by black circles in (a). At $k_3 = K_{YM}$, these bulk states satisfy the chiral symmetry. The eigenstates (g-h) $|\Phi\rangle$ located at $\omega_p + \sqrt{2n} \cdot v_\parallel \alpha$ have chiral symmetric partners (k-l) $\Gamma_3 \cdot |\Phi\rangle$ located at $\omega_p - \sqrt{2n} \cdot v_\parallel \alpha$. The horizontal/longitudinal modes in each chiral pair have the same/opposite phase distribution.

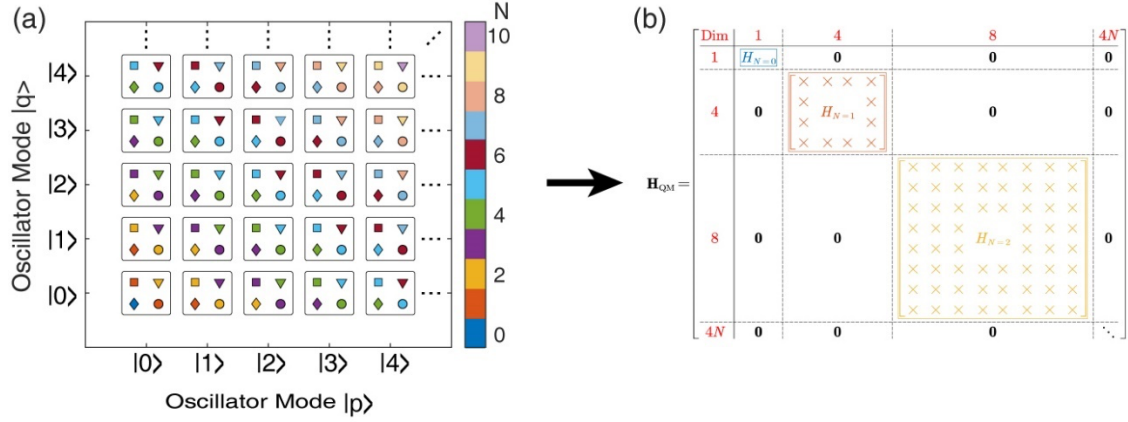

**Fig. S5. Schematic diagram of the blocked diagonal Hamiltonian $H_{QM}$.** (a) Distribution of the mode profiles in the virtual lattice, for the quantized supersymmetric Landau levels $\omega^2 = v_3^2 k_3 + 2v_\parallel^2 |B| \cdot N$ in the nonrelativistic squared Hamiltonian $H_{YM}^2$, with $B_{15} = B_{24} = B$ and $k_3 = 0$. The quantized energy is labeled by different colors. (b) Illustration of the block diagonal Hamiltonian $H_{QM}$. For a relativistic eigenstate at $\omega_C$, only those Fock states corresponding to $\omega_C^2$ in the non-relativistic squared Hamiltonian matter.

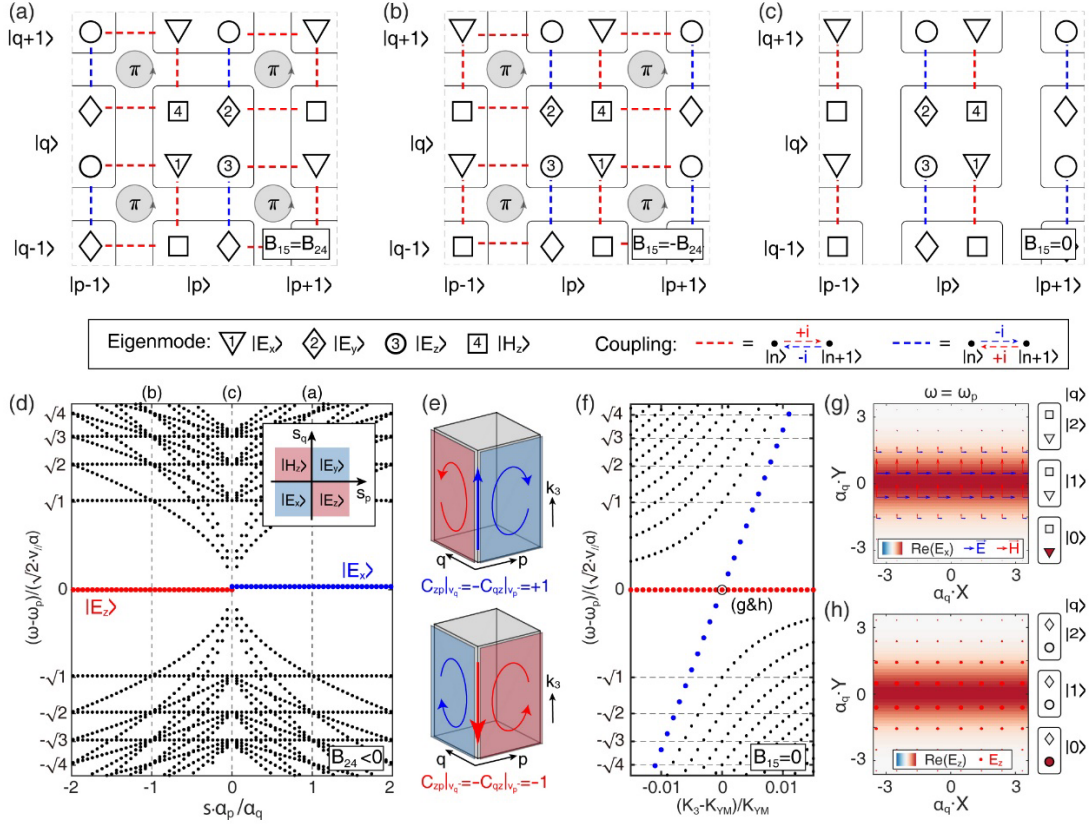

**Fig. S6. Schematic diagram of the periodic lattice model and the 1D limitation.** (a-c) Illustration of the flattened 2D periodic lattice model, with $B_{15} = B_{24}, -B_{24}$ and $0$, respectively. (d) The dispersion evolution for various $s \cdot \alpha_p/\alpha_q$ and a fixed $B_{24} < 0$, and with a very small $U_0$. The inset shows the eigenstate of the zeroth mode for the four different quadrants. (e) The existence of a hinge-localized mode due to the orthogonal but opposite two nested first Chern numbers defined on the Wannier sectors: $c_{zp}^{v_q^\pm} = -c_{qz}^{v_p^\pm} = \mp s = \mp s_p s_q$. (f) Dispersion spectrum with two helical zero modes at $B_{15} = 0$, where the 2D lattice model is converted to (c) a series of fully dimerized 1D SSH models. (g, h) The mode profiles in the virtual lattice and in real space for the helical zeroth modes in (f).

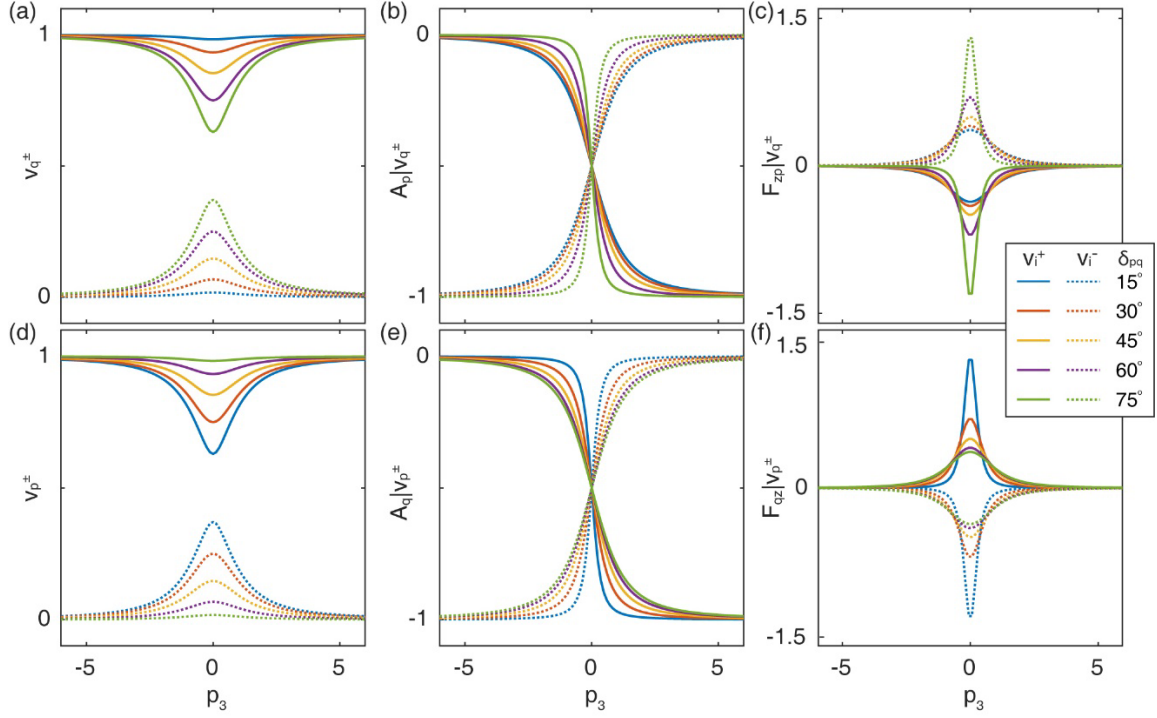

**Fig. S7. The evolution of the Wannier Berry connection and Berry Curvature.** (a) Construction of Wannier bands $v_q^\pm$ for the effective bulk Hamiltonian through Wilson loop diagonalization. (b, c) The evolution of (b) Berry connection and (c) Berry curvature defined on each Wannier sector. (d-f) The counterpart of (a-c), but for the Wannier bands $v_p^\pm$. In all these calculations, $s = s_p \cdot s_q = 1$, $\alpha_p = \cos(\delta_{pq})$ and $\alpha_q = \sin(\delta_{pq})$.

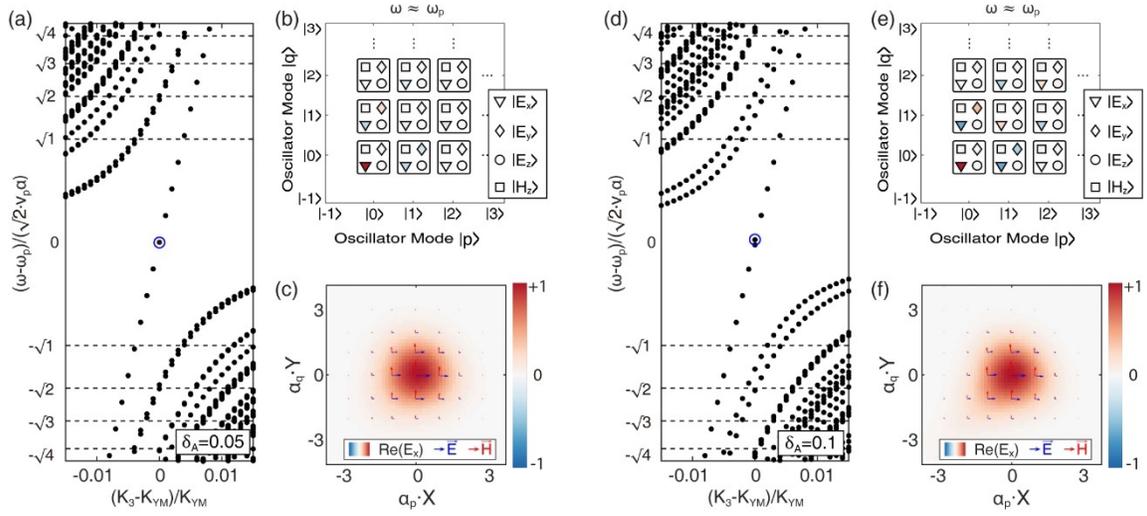

**Fig. S8. The robustness of the topologically protected CZM.** (a) The complete dispersion spectrum for the quasi-3D lattice model with an inhomogeneous background magnetic field, i.e., $\Delta_A = \delta_A \cdot B/L_0$, with $\delta_A = 0.05$. (b, c) The corresponding field distribution (b) in the virtual lattice and (c) in real space for the CZM (labeled by the blue circle). In the virtual lattice, only the first three Fock states are shown. (d-f) The counterparts of (a-c), with $\delta_A = 0.1$. The topologically protected CZM exists even when the background field is non-uniform. In these calculations, $B_{15} = B_{24} = B < 0$ and $L_0 = 10\lambda_p$.

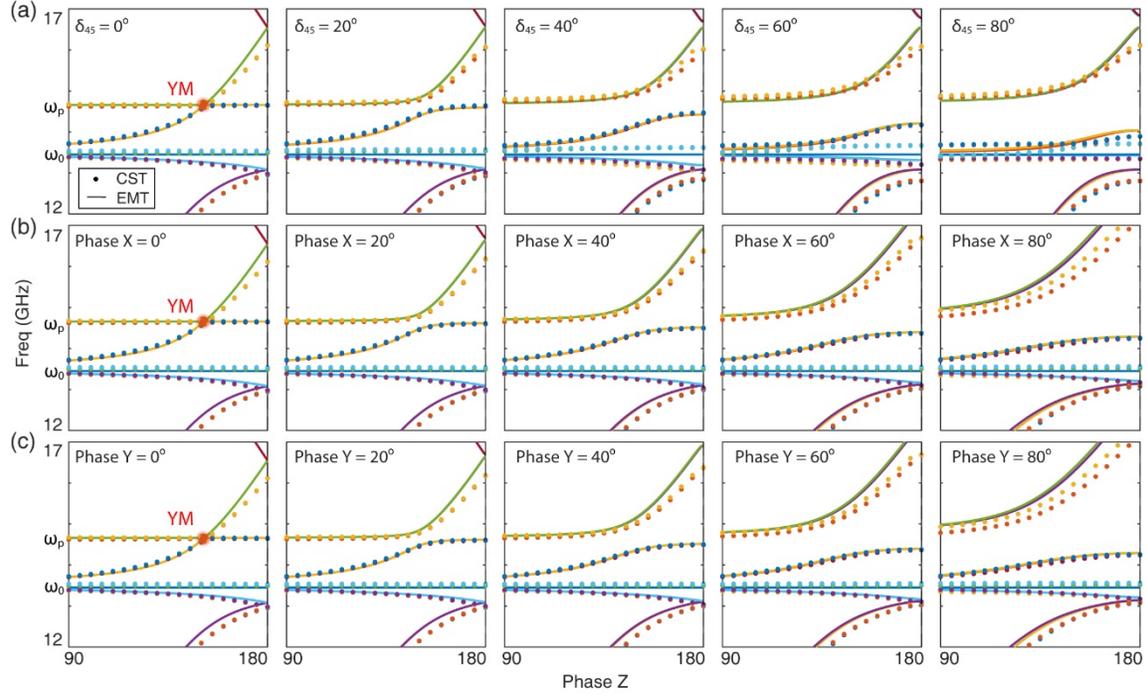

**Fig. S9. The Simulated and fitted dispersion of the designed Yang metamaterial.** (a) The evolution of dispersion along $z$ direction at different $\delta_{45}$ angles. (b, c) The counterpart of (a), but for different wavevectors (b) $k_x$ and (c) $k_y$. The dots and lines represent the simulated (CST) and Fitted (EMT) results, respectively. Near YM, $k_x$, $k_y$ and the synthetic wavevectors related to $\delta_{45}$ have almost the same effect on the dispersion spectrum.

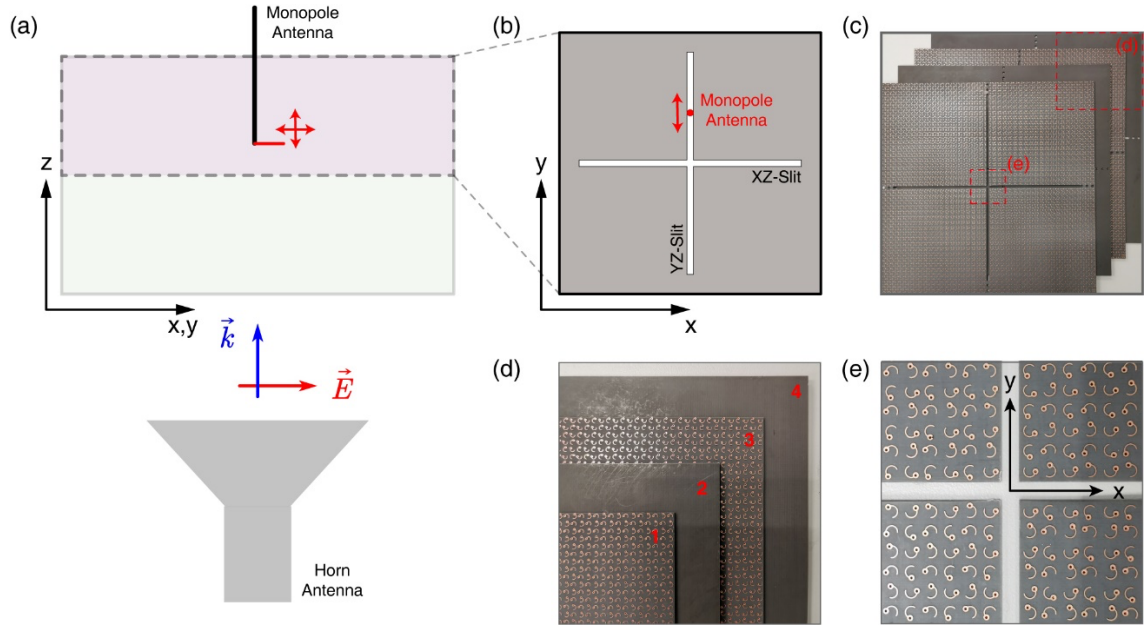

**Fig. S10. Schematic diagram of the experimental setup.** (a, b) Schematic of a co-polarization setup to detect the field distributions inside the metamaterial. (c) Photographs of the metamaterial with each unit cell containing four layers in *z* direction. (d, e) The zoomed photographs of different areas in (c).

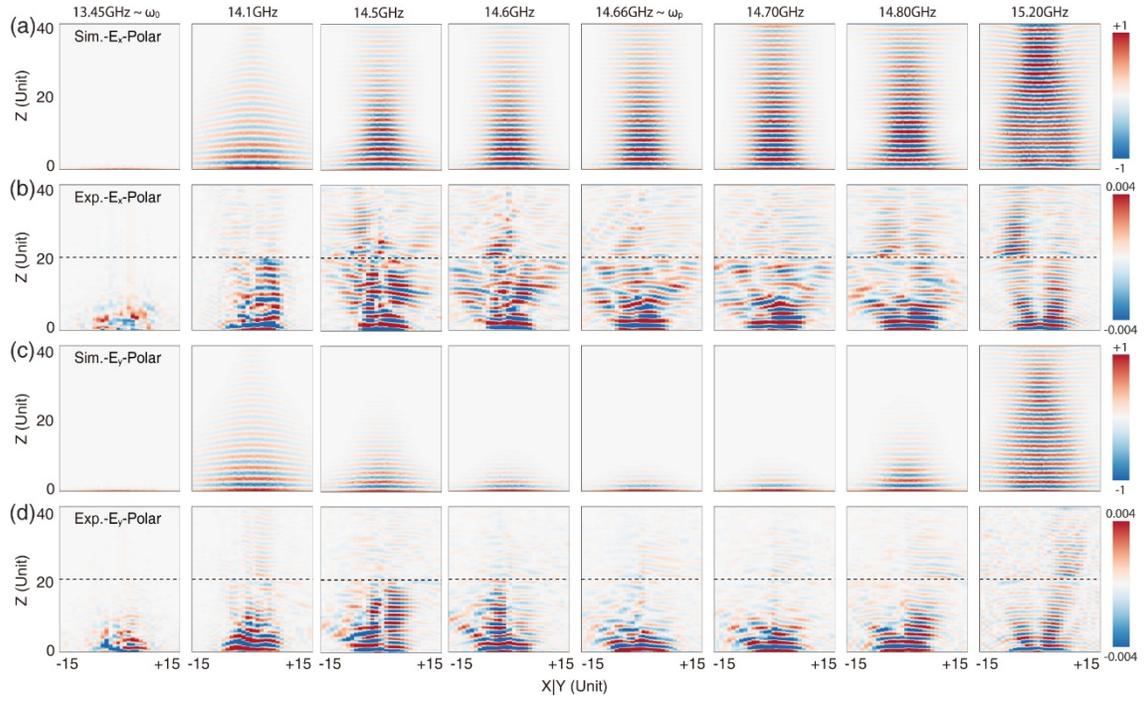

**Fig. S11. Polarization-dependent field distributions inside the metamaterial.** (a) The simulated $E_x$-field distribution inside the $x$-$z$ slit at different frequencies, excited by $E_x$-polarized incident wave. (b) The corresponding experimental measurement results. Since there are slits in only 80 layers (20 Units) of the samples, the data $z \in (0,20)$ units and $z \in (20,40)$ units are obtained from two independent measurements. (c, d) The same as (a, b), but for $E_y$-field distribution inside the $y$-$z$ slit, excited by $E_y$-polarized incident wave.

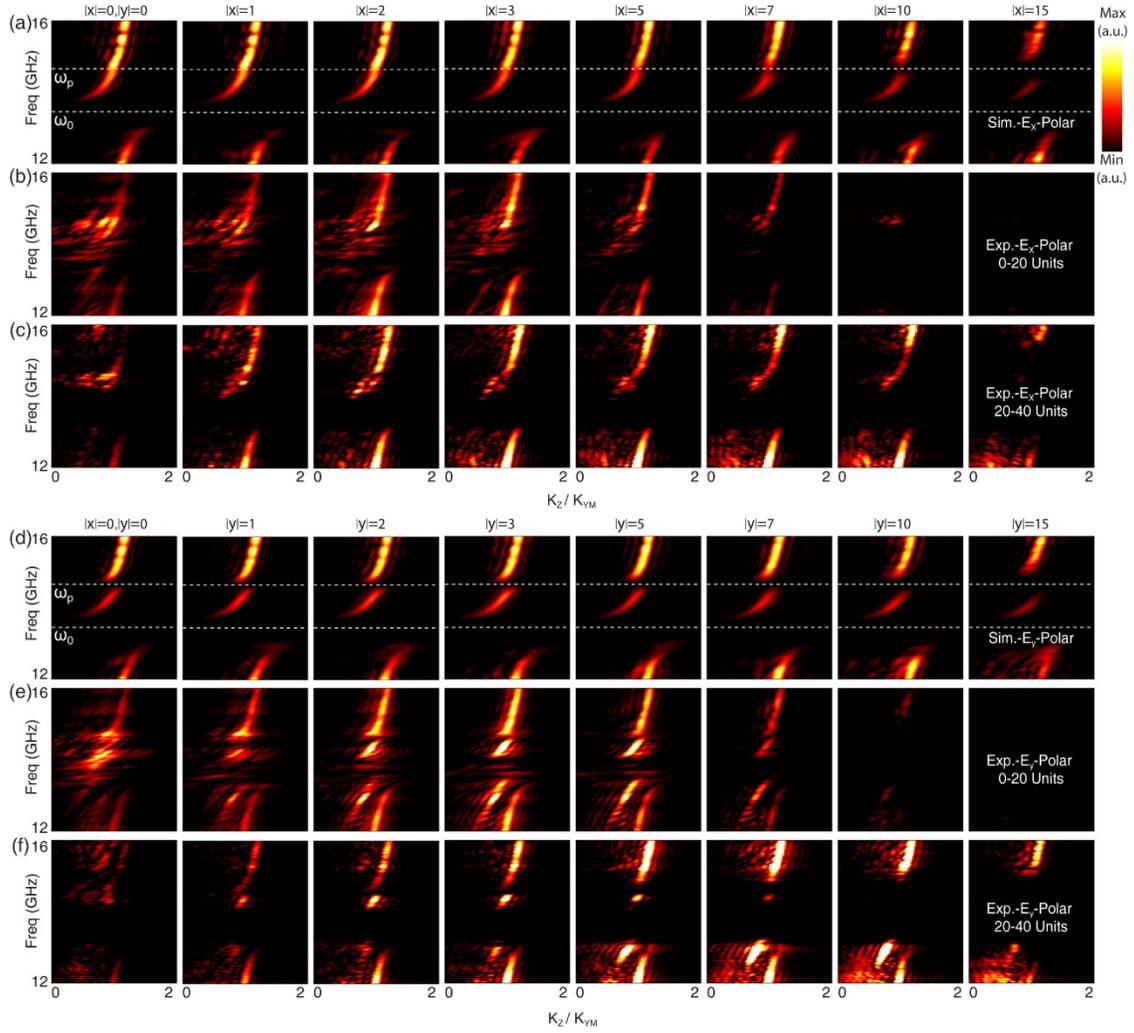

**Fig. S12. Experimental observation of CZM.** (a) The simulated dispersion spectra along $k_z$ with $E_x$-polarized incident wave, calculated by Fourier transformation of the $z$-dependent $E_x$-field distribution at a series of horizontal locations inside the $x$-$z$ slit. (b, c) The dispersion spectra obtained by two separate measurements in Fig. S11, (b) $z \in (0,20)$ units and (c) $z \in (20,40)$ units. (d-f) The counterpart of (a-c) for $E_y$-field distribution inside the $y$-$z$ slit excited by $E_y$-polarized incident wave. In each subpanel, the color bar is fixed, with an arbitrary unit from 0.1 to 1 to minimize the influence of the background for a better contrast. In addition, the $K_{YM}$ positions of the simulated/measured data are different, with $k_{YM} = 337.3 \text{ m}^{-1}$ in simulated model and $k_{YM} = 486.9 \text{ m}^{-1}$ in experimental measurement.

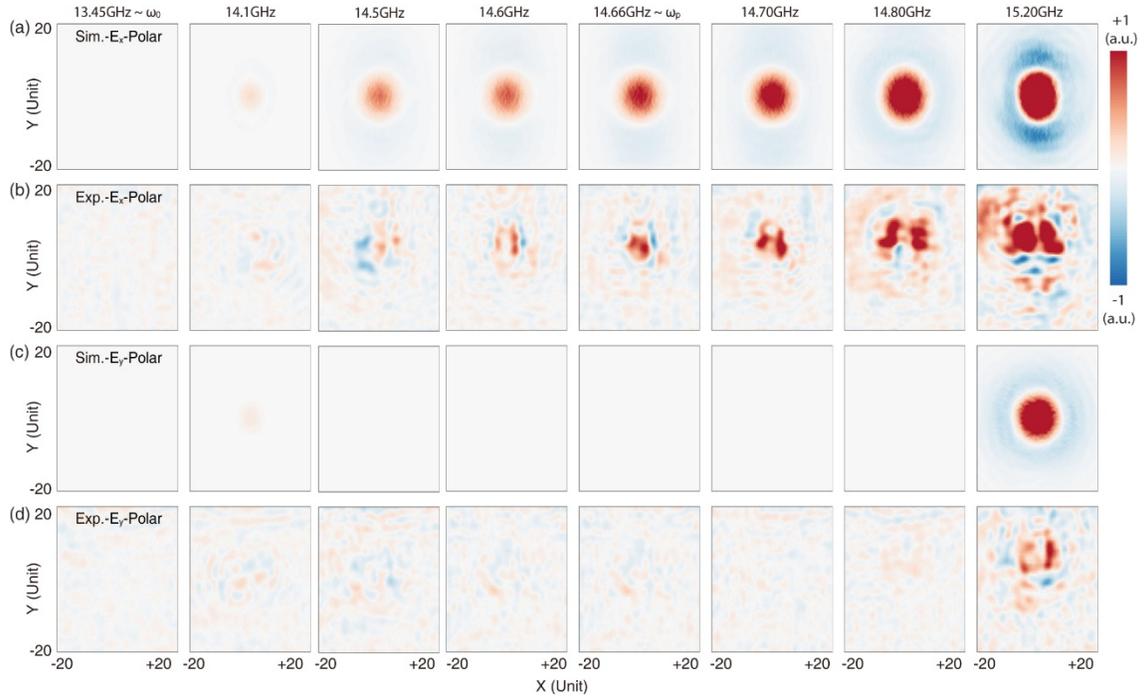

**Fig. S13. Simulated/measured polarization-dependent field distribution on the top surface of a 40-unit sample.** (a) The simulated $E_x$-field distribution in the *x-y* plane at different frequencies, with $E_x$ excitation. (b) The corresponding experimental results measured on a plane approximately 15 mm above the top surface. (c, d) The counterpart of (a, b) for $E_y$-field distribution, with $E_y$ excitation. The range of the color bar is fixed for simulated/experimental data, respectively.

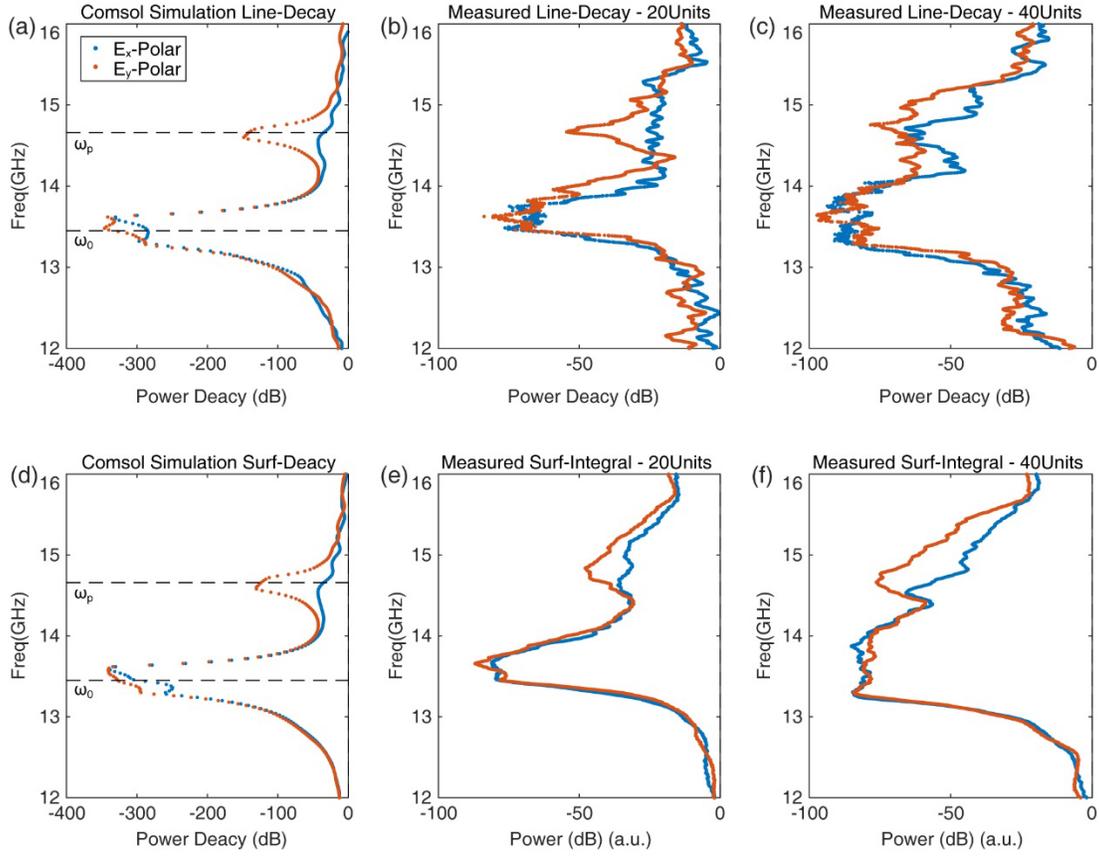

**Fig. S14. Simulated/measured polarization-dependent transmission power.** (a-c) The (a) simulated (20 Units) and (b, c) experimental transmission power for different polarizations, obtained by line integration of fields in Fig. S11. (d-f) The counterpart of (a-c), obtained by surface integration of fields in Fig. S13.